\documentclass[aip,graphicx,amsmath,amssymb,reprint]{revtex4-1}

\usepackage{graphicx}
\usepackage{amsmath}
\usepackage{amssymb}
\usepackage{dcolumn}
\usepackage{bm}
\usepackage[utf8]{inputenc}
\usepackage[T1]{fontenc}
\usepackage{mathptmx}
\usepackage{etoolbox}
\usepackage[symbol]{footmisc}
\usepackage{color}
\usepackage[dvipsnames]{xcolor}
\usepackage{hyperref}

\graphicspath{{Images/}}

\draft 

\begin{document}

\title{Data-driven approximations of topological insulator systems} 



\author{Justin T. Cole}
\author{Michael J. Nameika}
\affiliation{Department of Mathematics, University of Colorado,  1420 Austin Bluffs Parkway, Colorado Springs, 80918}


\date{\today}


\begin{abstract}
A data-driven  approach to calculating tight-binding models for discrete coupled-mode systems is presented. Specifically, spectral and topological data is used to build an appropriate discrete model that accurately replicates these properties. This work is motivated by topological insulator systems that are often described by  tight-binding models. The problem is formulated as the minimization of an appropriate  residual (objective) function.
Given bulk spectral data and a topological index (e.g. winding number), an appropriate discrete  model is obtained to arbitrary precision. 
A nonlinear least squares method  is used to determine the coefficients. 
The effectiveness of the scheme is highlighted against a Schr\"odinger equation with a periodic potential that can be described by the Su-Schrieffer-Heeger  model.
\end{abstract}

\maketitle 


\section{Introduction}
A central challenge across all fields of science is the derivation of appropriate mathematical models. Ideally, these  are simple (solvable) equations that capture all essential behavior.
One approach that is currently receiving significant interest is that of data-driven machine learning methods for the derivation of governing equations, e.g. differential equations \cite{Brunton2016,Rudy2017}. The problem setup is: given a set of relevant measurements e.g. field data at a given space and/or time, find the terms of the governing equation, e.g. coefficients.

This work focuses on deriving models for a class of systems known as topological insulators \cite{Hasan2010,Qi2011}. These systems are characterized by localized  states that are exceptionally robust to 
defects. Moreover, these modes are said to be topologically protected, as they possess   topological invariants that are associated with the localized states \cite{Thouless1982}. The remarkably stable nature of these states makes them a good candidate for applications such as isolators \cite{Jalas2013}, lasers \cite{Bahari2017}, or switches \cite{Ablowitz2024}.

A popular class of models for understanding topological insulators is that of tight-binding models \cite{Ablowitz2022,Kevrekidis2009}. These are discrete models that describe  interactions among localized modes, e.g. Wannier functions \cite{Alfimov2002}, in lattice systems. The exponential decay of these functions motivates the truncation of nonlocal modal  interactions. Hence, from variable coefficient differential equations, one can derive constant coefficient difference equations that 
considerably simplify the analysis of the system, yet still capture the dominant behaviors.

Two popular approaches  for deriving tight-binding models is (a) maximally localized Wannier expansion \cite{Marzari1997} and (b) linear combination of atomic orbitals \cite{Bloch1929}. These are direct approaches. A Bloch wave function is expanded in terms of Wannier or orbital functions and from it a reduced order model is derived in a Galerkin manner.  These are well-established and thoroughly studied approaches for building tight-binding models \cite{Ablowitz2012,Fefferman2018}.

The  goal of this work is to develop an {\it indirect} approach for establishing tight-binding models that describe topological insulators systems. From a set of user-provided spectral bands,  an optimization algorithm is used to fit the coefficients of a tight-binding model.
These types of ideas are not new and can be traced back to the work 	of Slater and Koster \cite{Slater1954,Papaconstantopoulos2003}. 
That work highlighted several assumptions, e.g. symmetries and number of modes, to reduce the complexity of the problem.

A central challenge here is the non-uniqueness of the bulk spectral data. As an example, for the Su-Schrieffer-Heeger (SSH) model both the topologically trivial and nontrivial systems can possess exactly the same bulk spectral bands; only the eigenfunctions reveal their topological nature. It will be important to design methods that find the model with the appropriate topology.

In this work a nonlinear least squares algorithm is applied to the fitting of tight-binding coefficients. The objective function to be minimized is the (residual) difference between given bulk spectral data and the discrete approximation. A Levenberg-Marquardt  method is used to minimize the error and find an optimal choice of parameters. The optimization process is automated. All gradients are computed symbolically, that is exactly, using  MATLAB's symbolic computation toolbox. 

Due to the automatic nature of our approach, we are able to incorporate an arbitrary number of interactions with relative ease. That is, the user may input the desired number of (nonlocal) interactions beyond nearest neighbor, typically at the cost of longer run times. Hence, for most lattices considered, we are able to find 
tight-binding models of arbitrary precision. On the other hand, we find that additional interactions must be added in such a way to preserve the symmetries of the model, e.g. inversion symmetry.

As an example, we apply this approach to the SSH model. This is a prototypical topological insulator model that describes the dynamics of a one-dimensional array of identical molecules who couple to their two adjacent neighbors. 
Originally proposed in the context of hydrocarbon chains \cite{Su1979}, it has subsequently been applied in numerous applications, such as photonic waveguide arrays \cite{Malkova2009,Xia2021,Weimann2016,Ivanov2023}, Bose-Einstein condensates  in optical lattices \cite{Atala2013}, and varying cross-section acoustic cavities \cite{Coutant2021}, among others. The unifying ingredient in all these systems is an alternating set of hopping coefficients which can induce localized edge states (see Sec.~\ref{SSH_review_sec}). 


One of the benefits of this approach is the models can be derived directly from data. That is, it is not necessary to first solve a differential equation to obtain the spectral data. Instead, this data could perhaps be derived from experimental measurements, or some other first-principles calculation. Hence, the name {\it data-driven}. In this work, we obtain our data from a linear Schr\"odinger equation with a periodic potential. In the deep lattice limit, the SSH model is the effective tight-binding model.


We now give the outline for this paper. In Section~\ref{Schrodinger_sec} we review the basic equations of 
this work. The Schr\"odinger equation and the Zak phase are introduced and some relevant properties reviewed. In Section~\ref{wannier_func_sec} we review the Wannier functions and the derivation of tight-binding models in the context of the one-dimensional Schr\"odinger equation. The SSH model and some associated properties are reviewed in Appendix~\ref{SSH_review_sec}. In Section~\ref{optimize_sec} we introduce our data-driven optimization approach for deriving tight-binding models that describe topological insulators. Typical results for the method are shown in Section~\ref{results}. Some connections to the coefficients obtained through the maximally localized Wannier function approach are shown in Section~\ref{wannier_comparison}. We conclude in Section~\ref{conclude_sec}.


\section{The Schr{\"o}dinger Equation}
\label{Schrodinger_sec}

The  Schr\"odinger equation is a prototypical differential equation for describing a wide variety of physical systems, such as quantum mechanics \cite{Schrodinger1926,Griffiths2018}, Bose-Einstein condensates \cite{Gross1961,Pitaeveskii1961}, paraxial electromagnetic beams \cite{Kivshar2003,Ablowitz2011}, and deep water waves \cite{Zakharov1968}, among others. 
Consider the one dimensional (1D) time-independent Schr{\"o}dinger equation
\begin{equation}
\label{Schrodinger_eqn}
    \frac{d^2\psi}{d x^2} + V(x)\psi = \mu \psi ,
\end{equation}
with eigenvalue or propagation constant $\mu$ and corresponding eigenfunction  $\psi$. The wavefunction is  subject to diffractive/dispersive Laplacian effects and a trapping potential modeled by $V(x)$. In the case of photonic lattices, the potential is proportional to the local index of refraction contrast from the background \cite{Kivshar2003,Ablowitz2011}. 

Consider real periodic potentials with period $L > 0$, such that $V(x + L) = V(x)$. 
 Bloch theory \cite{Kuchment1982} motivates Bloch wave solutions of the form
\begin{equation}
\label{bloch_wave}
    \psi(x,k) = e^{ikx}u(x,k) , ~~~~~~~~~~~ u(x + L, k) = u(x,k) ,
\end{equation}
with quasi-periodic boundary conditions $\psi(x + L , k) = e^{ i k L} \psi(x,k) $ for quasimomentum $k$. 
We are interested in bounded Bloch solutions (\ref{bloch_wave}), so we restrict our attention to real values of $k$ in the Brillouin zone $[- \frac{\pi}{L} , \frac{\pi}{L}]$. Then (\ref{Schrodinger_eqn}) becomes
\begin{equation}
\label{bloch_wave_eqn}
    \frac{d^2u}{d x^2} + 2ik\frac{d u}{d x} - k^2u + V(x)u = \mu(k)u ,
\end{equation}
where $\mu(k)$ indicates a spectral band. 
Since the potential is assumed to be real, i.e. $V^*(x) = V(x)$, one can show that  
the  operator in (\ref{bloch_wave_eqn}) 
is Hermitian and so the spectrum is real, i.e. $\mu^*(k) = \mu(k)$, for all $k$. Moreover, the system also possess time-reversal symmetry and the spectral bands exhibit the inversion symmetry, $\mu(-k) = \mu (k)$.

We focus our attention potentials with inversion symmetry
\begin{equation}
\label{inversion_symm}
V(-x) = V(x),  ~~~~~ V(L - x)  = V(x) ,
\end{equation}
which possess inversion centers at $x = 0 , L/2$ (up to the lattice period $L$). This is a natural potential to consider when the aim is to derive an SSH-type tight-binding model. Explicitly, we consider the (periodic) sum of Gaussians
\begin{equation}
    \label{pot_expand}
    V(x) = \sum_{n = -\infty}^{\infty} V_0^2\left[e^{-\frac{(x - nL - x_1)^2}{\sigma^2}} + e^{-\frac{(x - nL + x_1)^2}{\sigma^2}}\right] , 
\end{equation}
where $V_0^2$ is the potential strength or depth, $x_1$ corresponds to so-called $a$-sites and $-x_1$ corresponds to $b$-sites for $x_1 < 0$. 
Observe that this potential is real, $L$-periodic, and satisfies the inversion symmetry (\ref{inversion_symm}).
The value $\sigma > 0 $ is a  tuning parameter for the Gaussian widths. 


To numerically approximate the eigenvalue and eigenfunctions of (\ref{bloch_wave_eqn}), we employ a Fourier collocation method \cite{Trefethen2000,Fornberg1998}.
The periodic domain $\left[- \frac{L}{2} , \frac{L}{2} \right]$ is discretized by a uniformly spaced grid.
The derivatives are approximated by Fourier spectral differentiation matrices (see \cite{Weideman2000}).
At a fixed $k$ value, the discretized version of (\ref{bloch_wave_eqn}) is solved. The eigenvalue-eigenfunction pair $\{ \mu_p(k) , u_p(x,k) \}$ for the $p^{\rm th}$ spectral band are numerically computed in a standard eigenvalue solver. In this work, we only consider the first two bands, sorted in descending order.  Denote the first and second bands by $\mu_1(k)$ and $\mu_2(k)$, respectively.

\begin{figure}
   \includegraphics[scale = 0.19]{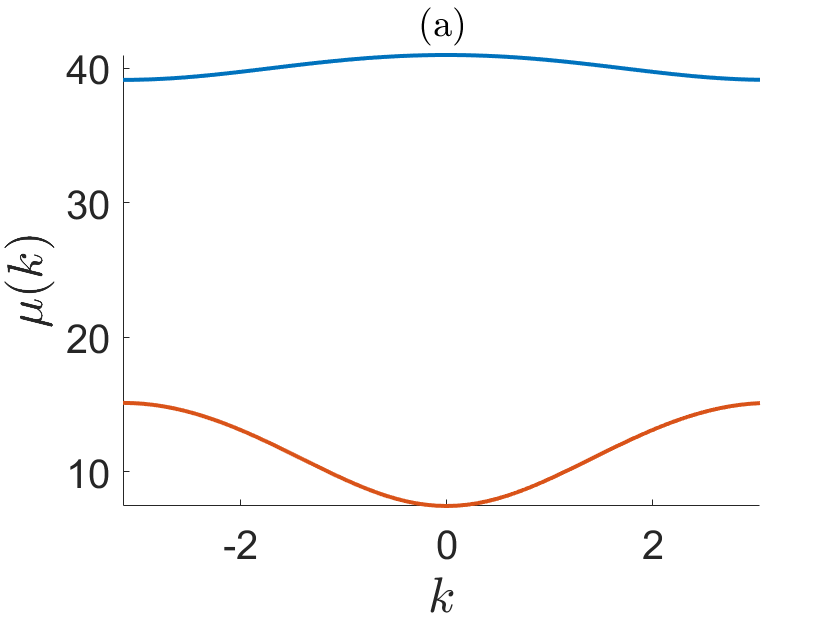}
    \includegraphics[scale = 0.19]{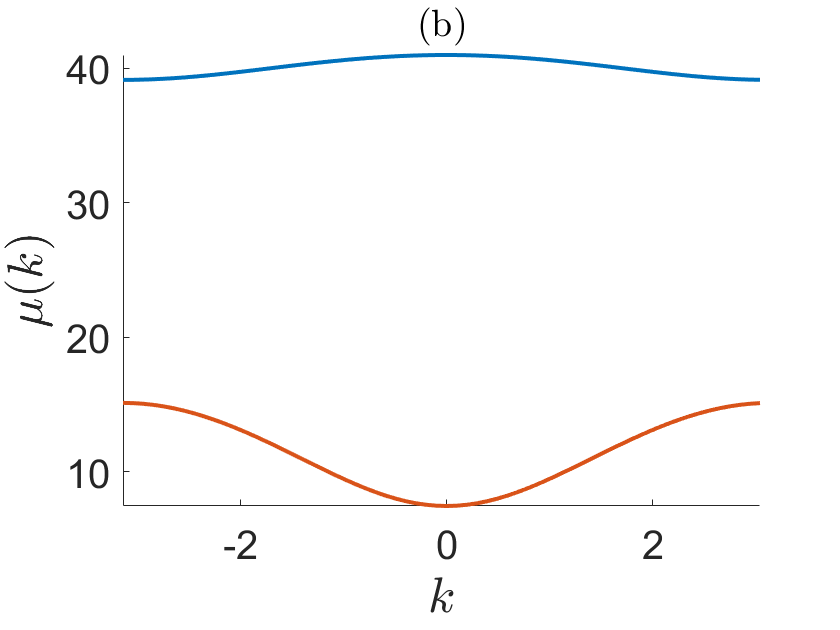}
    \newline
    \includegraphics[scale = 0.19]{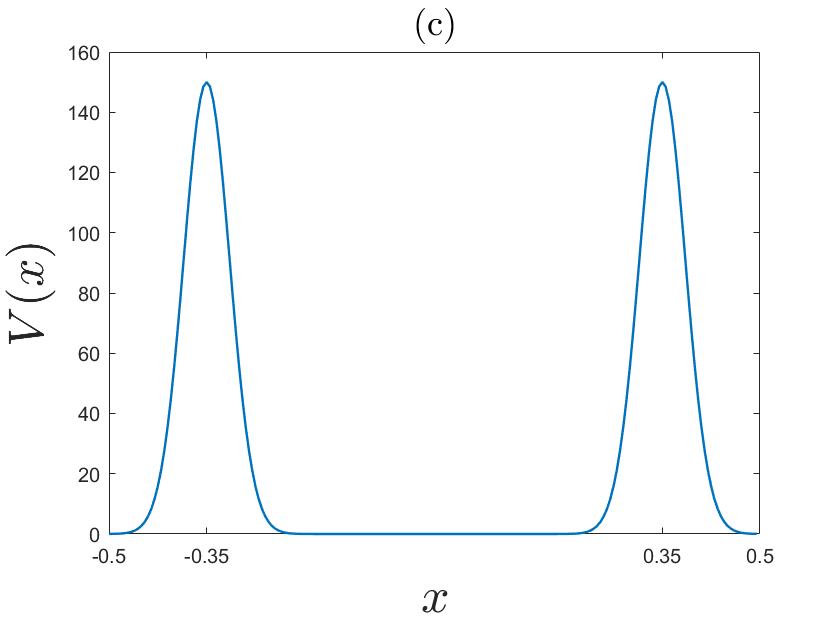}
    \includegraphics[scale = 0.19]{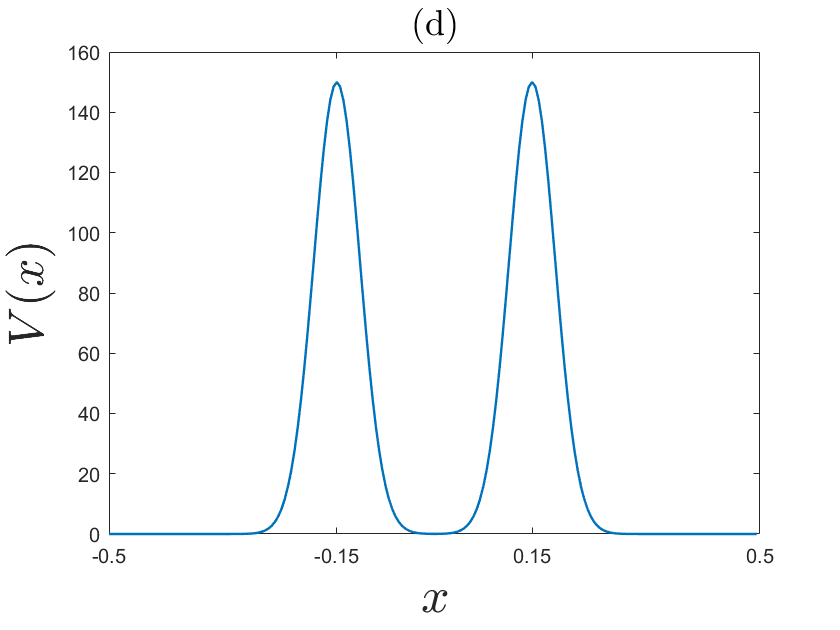}
    \centering
    \caption{Top row: Spectral bands, sorted in descending order, generated by solving the Schr{\"o}dinger equation (\ref{bloch_wave_eqn}) with lattice period $L = 1$. Notice the bands are identical. Bottom row: Corresponding potentials (\ref{pot_expand}) with parameters 
(a,c) : $\sigma = 0.05$, $x_1 = -0.35$, $V_0^2 = 150$. 
 (b,d) : $\sigma = 0.05$, $x_1 = -0.15$, $V_0^2 = 150$. 
 \label{Schrodinger_bands_plot}}
\end{figure}

A typical set of potentials and their corresponding bands 
are shown in Fig.~\ref{Schrodinger_bands_plot} for one period of the sum of Gaussians given in (\ref{pot_expand}). Two distinct potential configurations are considered (see Fig.~\ref{Schrodinger_bands_plot}, bottom row). In each case considered, the first two spectral bands are shown and there is a gap between them. A gap remains open as long as $x_1 \not= -L/4$, 
in which case the smallest period of the lattice is $L/2$, not $L$.

Notice that the peaks in the  potential of Fig.~\ref{Schrodinger_bands_plot}(c) are a distance $0.15L$ from the boundaries, while in the  case shown in Fig.~\ref{Schrodinger_bands_plot}(d) the peaks are a distance $0.15L$ from the origin. These two potentials are observed to exhibit identical band configurations. This  is found to occur whenever peak distance from the origin is the {\it same} as from the spatial cell boundaries (at $x = \pm L/2$). These two lattice configurations turn out to be topologically distinct. The next section introduces the topological quantity to distinguish between them. 

\subsection{The Zak (Berry) phase}

Topological insulators  support modes associated with certain topological invariants. Here, the appropriate topological quantity is the Berry or Zak phase \cite{Berry1984,Zak1989}. Intuitively, this quantity detects intrinsic geometric phases of eigenmodes. The most obvious difference between different topologies is the presence of edge modes, or lack thereof.

The Zak phase is defined by 
\begin{equation}
\label{Zak_phase_cont}
\mathcal{Z} =  \oint_C  \mathcal{A}(k) dk , ~~~~ \mathcal{A}(k) = i \left\langle   u (x,k) \bigg| \frac{d u }{d k}(x,k) \right\rangle ,
\end{equation}
where $\mathcal{A}(k)$ is the so-called Berry connection for an appropriate inner product.
 This measures the amount of times the geometric phase winds as $k$ moves from $0$ to $2\pi / L$. 

The Zak phase (\ref{Zak_phase_cont}) is gauge variant and is only known up to an integer multiple of $2 \pi$. Also,  the Berry connection $\mathcal{A}(k)$ is periodic in $k$ due to the quasi-periodic boundary conditions of $u(x,k)$.
The numerical approximation of the Zak phase is discussed in Appendix~\ref{Zak_appendix}.
As a result of the inversion symmetry (\ref{inversion_symm}), the Zak phase is quantized\cite{Zak1989} and the topology is categorized via
\begin{equation}
    \label{winding_relation}
    \mathcal{Z} \equiv \begin{cases}
        0  & \Leftrightarrow ~~ \text{trivial topology}\\
        \pi  & \Leftrightarrow ~~ \text{nontrivial topology}
    \end{cases} .
\end{equation}

One immediate consequence of a nontrivial topology is the presence of edge modes. This relationship between the bulk topological invariant and the presence of edge states is known as the {\it bulk-edge correspondence}. Since the bulk problem does not see the edge, this invariant must be robust. Small defects and perturbations to the boundaries do not eliminate the presence of these edge modes.

\section{Wannier Functions}
\label{wannier_func_sec}

Before moving to the tight-binding approximation (the main focus of this work), some background on Wannier functions is introduced. Understanding these functions is important as they serve as a basis in which to expand the Bloch wave eigenfunctions of (\ref{Schrodinger_eqn}). Ultimately, however, we wish to bypass this {\it direct} approach for an {\it indirect} one.

To motivative the direct approach, note that the Bloch wave (\ref{bloch_wave}) is periodic with respect to the quasimomentum, i.e. $\psi \left(x,k + \frac{2 \pi }{L} \right) = \psi (x, k)$. As such, the Bloch wave eigenfunction corresponding to the $p^{\rm th}$ band 
can be expressed in the Fourier series
\begin{equation}
    \label{Bloch_Fourier}
    \psi_p(x,k) = \sum_{n = -\infty}^{\infty} {W}_{np}(x) e^{ i n k L} ,
\end{equation}
where the $p^{\rm th}$ band Fourier coefficients $W_{np}(x)$ are the  Wannier functions at lattice cell $n$, defined by
\begin{equation}
    \label{Wannier_coeff_compute}
    {W}_{np}(x) = \frac{L}{2 \pi} \int_{-\pi/L}^{\pi/L} e^{ -i n k L}  \psi_p(x,k) d k .
\end{equation}
When the Bloch function is infinitely differentiable in $k$, these Wannier functions decay to zero faster than any algebraic function\cite{Kohn1959}.  Wannier functions also possess the translation property
$    {W}_{0p}(x - n L ) = {W}_{np}(x) . $
 Said differently, all same-band Wannier functions are translations of each other. 
 Next, define the inner product
\begin{equation*}
    \langle f | g \rangle = \int_{-\infty}^{\infty} f^*(x)g(x)dx.
\end{equation*}
Wannier functions form an orthogonal basis with respect to this inner product. That is, $\langle W_{n',p'} | W_{n,p} \rangle = \delta_{n,n'} \delta_{p,p'}$. 

The value of the Zak (Berry) phase indicates the location (first moment) of the Wannier centers \cite{Blount1962}. That is, 
\begin{equation}
\label{mass_center_Wannier}
\overline{x}_p = \langle {W}_{0p}(x) | x {W}_{0p}(x) \rangle = \frac{L}{2\pi} \oint_C i \left\langle   {u} (k) \bigg\rvert \frac{d {u} }{d k} \right\rangle dk = \frac{L}{2 \pi} \mathcal{Z} ,
\end{equation}
where ${W}_{0p}(x)$ denotes the Wannier functions corresponding to the unit cell ($n = 0$) and the Zak phase is defined in (\ref{Zak_phase_cont}). 
Note that the inner product on the left hand side (over $\mathbb{R}$) is different from the one on the right side (over unit cell $[-L/2 , L/2]$). 
From the topological values in (\ref{winding_relation}), we observe that a null (nonzero) Zak phase indicates a Wannier center located in the center (edge) of the unit cell. Concatenating a finite number of period cells for the nonzero Zak phase leads to an isolated Wanner function located at the endpoints of the array. As a result, we expect to find a localized so-called {\it edge mode} that decays rapidly into the lattice, away from the edges. This is an instance of the {\it bulk-edge correspondence} principle that relates the bulk topological invariants (i.e. Zak phase) to the presence of edge states in the boundary value problem.

\begin{figure}
    \includegraphics[scale = 0.19]{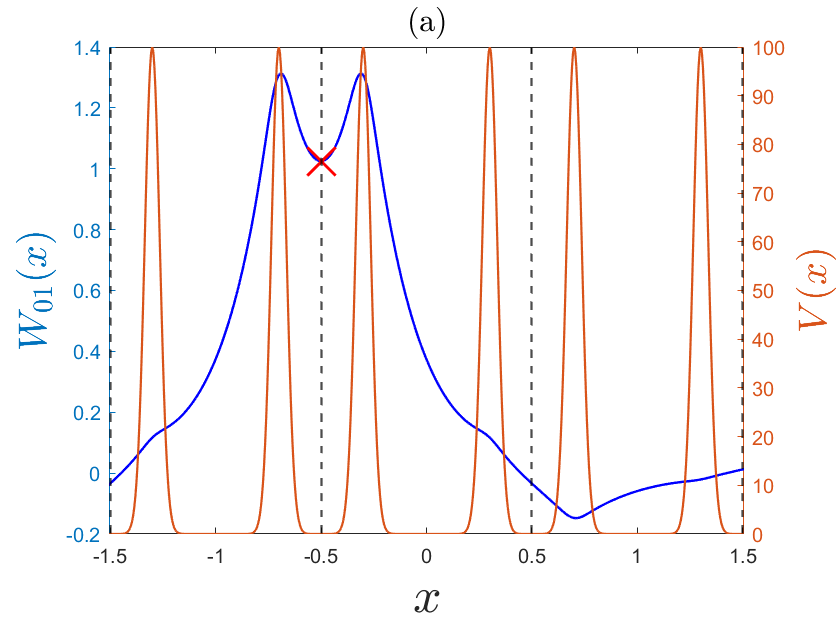}
    \includegraphics[scale = 0.19]{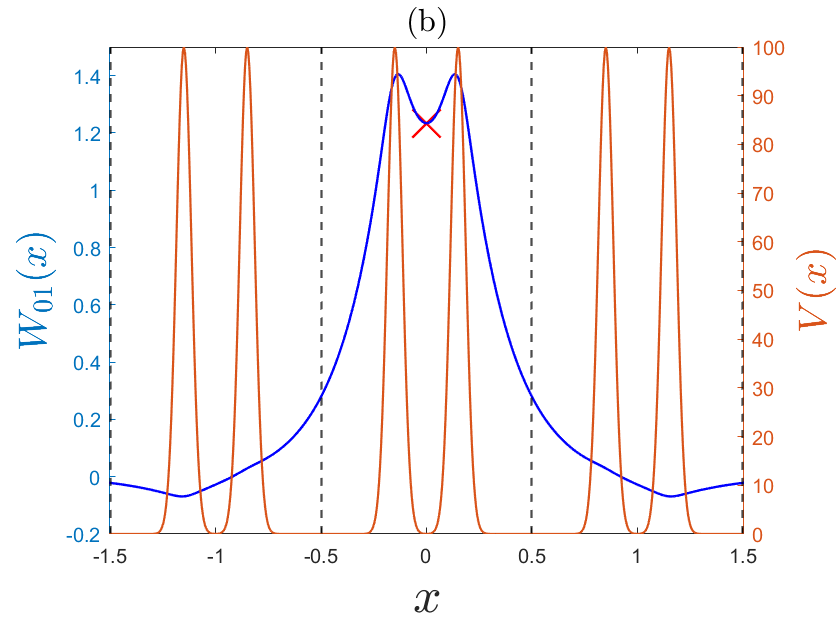}
    \newline
    \includegraphics[scale = 0.19]{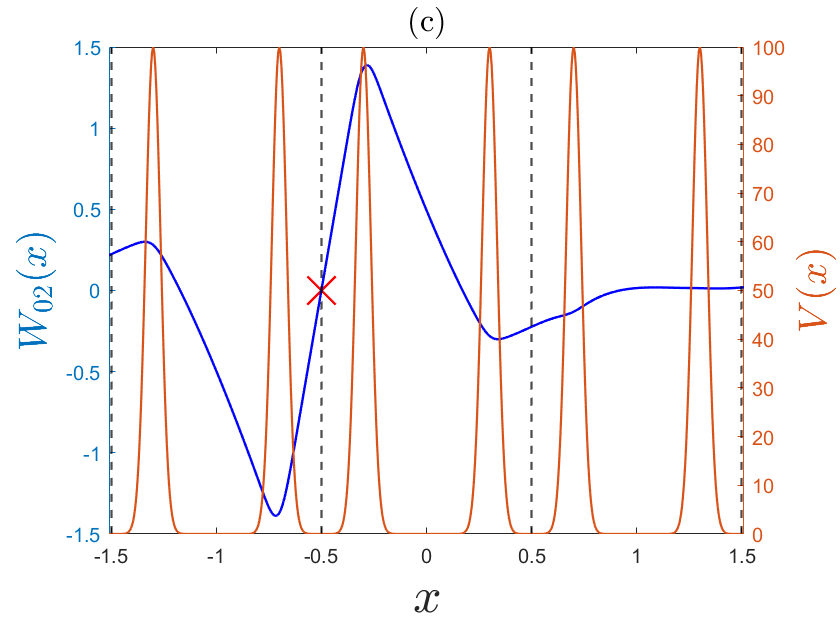}
    \includegraphics[scale = 0.19]{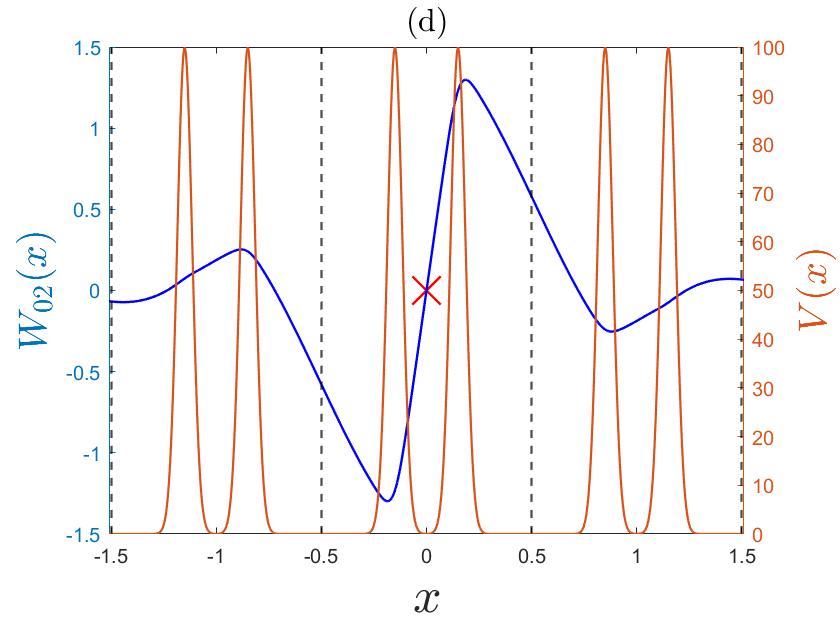}
    \centering
    \caption{Numerically computed MLWFs for the first two bands using potential (\ref{pot_expand}) plotted against the potential used to create them. For each case the potential parameters are  $L = 1, V_0^2 = 100, \sigma = 0.05$. The left (right) column depicts a topological 
  (nontopological) potential. The topological potential (panels (a) and (c))  has peaks located at $x = -0.3 + n$ and $x = 0.3 + n$, for $n \in \mathbb{Z}$. The nontopological potential (panels (b) and (d)) has peaks located at $x = -0.15 + n$ and $x = 0.15 + n$.  Dashed lines depict unit cell boundaries. Red `x' marks the Wannier function center of mass (first moment).  \label{MLWF_plot}}
\end{figure}

Using the $PXP$ operator (see Appendix~\ref{num_calc_wannier_modes}), a typical set of Wannier functions (\ref{Wannier_coeff_compute}) for a single band is shown in Fig.~\ref{MLWF_plot} corresponding to potential (\ref{pot_expand}). 
 All Wannier modes shown are real and exponentially localized. Moreover, in each case the maximum amplitude 
 values occur at the potential centers. This is expected since these potential regions act as dielectric (attractive) waveguides.
In addition, each mode shown is not sign-definite
; the tails oscillate.
This changing of sign is essential for achieving orthogonality with real functions. 

From these Wannier functions we may  infer the Zak phase using (\ref{mass_center_Wannier}). In the topologically nontrivial or simply ``topological'' configuration (see Figs.~\ref{MLWF_plot}(a,c)), there is a Zak phase (\ref{Zak_phase_cont}) of $\mathcal{Z} = \pi$ (up to integer shifts of $2 \pi$) and so the Wannier center (\ref{mass_center_Wannier}) is located at the cell edge $x = -L/2$. On the other hand, the topologically trivial or ``nontopological'' configuration with $\mathcal{Z} = 0$  (see Figs.~\ref{MLWF_plot}(b,d)) is centered at the origin (up to an integer multiple of $L$ shift).  One can view these Wannier functions as being intercell and intracell dominated, respectively.

\subsection{A Wannier-Galerkin tight-binding approximation}
\label{Wannier_Galerkin}

In general, numerics are required to solve differential equations with periodic coefficients. However, it is desirable to obtain some analytical results, even if it requires some simplifying assumptions.
For linear differential equations with periodic coefficients, like (\ref{Schrodinger_eqn}), an effective way of obtaining reduced order models is through the tight-binding approximation. Using this approach, one can reduce a differential equation with variable coefficients to an algebraic system of difference equations with constant coefficients.  This approach is effectively a Galerkin method \cite{Canuto2006}. That is, the solution is expanded in terms of an appropriate basis and the system is recast in a weak (integral) formulation.

Recast the Fourier series (\ref{Bloch_Fourier}) in the expansion of Wannier-type modes 
\begin{equation}
    \label{two_band_wannier_expand}
    \psi(x) = \sum_{n = -\infty}^{\infty} \left[ a_n {W}_{n,1}(x) + b_n {W}_{n,2} (x) \right].
\end{equation}
where $a_{n} (b_n)$ denotes the modal coefficient corresponding to the $n^{\rm}$th spatial cell and the Wannier mode centered at $x_1 (-x_1)$; cf. (\ref{pot_expand}). 
Since we will be working with a two band model, we restrict our attention to the first two bands. A typical set of single-band Wannier functions is shown in Fig.~\ref{MLWF_plot} where all Wannier-type functions are assumed to be real and exponentially localized. On the other hand, the maximally localized Wannier functions represent a linear combination of these modes. 
Within the $n^{\rm th}$ cell, $a_n$ and $b_n$ correspond to the fundamental and secondary modal coefficients, respectively.

Next, substitute  expansion (\ref{two_band_wannier_expand})  into (\ref{Schrodinger_eqn}), multiply by a test function ${W}_{m,p}$, and integrate to obtain
\begin{widetext}
    \begin{eqnarray}
    \label{wannier_psi_a}
    \sum_{n = -\infty}^{\infty} a_n[   \langle {W}_{m,1} | V(x) {W}_{n,1} \rangle - \langle {W}'_{m,1} | {W}'_{n,1} \rangle] + b_n[  \langle {W}_{m,1} | V(x){W}_{n,2}\rangle - \langle {W}'_{m,1} | {W}'_{n,2}\rangle] 
      & = \mu  a_m  , \\
\label{wannier_psi_b}
    \sum_{n = -\infty}^{\infty} a_n[   \langle {W}_{m,2} | V(x) {W}_{n,1} \rangle- \langle {W}'_{m,2} | {W}'_{n,1} \rangle] + b_n[  \langle {W}_{m,2} | V(x){W}_{n,2}\rangle - \langle {W}'_{m,2} | {W}'_{n,2}\rangle] 
      & = \mu  b_m .             
    \end{eqnarray}
\end{widetext}
for $ p =1,2$, respectively. Note that the orthogonality of the Wannier functions and an integration-by-parts has been applied.

In the tight-binding limit the Wannier functions decay super-algebraically fast. That is, for $p,q = 1,2$, it follows that $ \langle {W}_{m,p} | V(x) {W}_{n,q} \rangle, \langle {W}'_{m,p} |  {W}'_{n,q} \rangle \rightarrow 0$ fast as $|n-m| \rightarrow \infty$. As such, we truncate  nonlocal terms, which are typically quite small, and keep only nearby sites.
The resulting system of equations is
\begin{widetext}
    \begin{eqnarray}
    \label{wannier_psi_a2}
    \sum_{n = m - \text{dist} +1}^{m+{\text{dist}}-1} a_n[   \langle {W}_{m,1} | V(x) {W}_{n,1} \rangle - \langle {W}'_{m,1} | {W}'_{n,1} \rangle] +  \sum_{n = m -  \text{dist}}^{m+ \text{dist} -1} b_n[  \langle {W}_{m,1} | V(x) {W}_{n,2}\rangle - \langle {W}'_{m,1} | {W}'_{n,2}\rangle] 
      && = \lambda  a_m  , \\
\label{wannier_psi_b2}
    \sum_{n = m -  \text{dist} + 1}^{m+{\text{dist}}} a_n[   \langle {W}_{m,2} | V(x){W}_{n,1} \rangle- \langle {W}'_{m,2} | {W}'_{n,1} \rangle] +  \sum_{n = m -  \text{dist} + 1}^{m+ \text{dist} -1} b_n[  \langle {W}_{m,2} | V(x) {W}_{n,2}\rangle - \langle {W}'_{m,2} | {W}'_{n,2}\rangle] 
      && = \lambda  b_m ,      
    \end{eqnarray}
\end{widetext}
where $ \text{dist}  \in \{1,2,3, \dots \}$ is the distance of interactions considered. 
For example, $ \text{dist} = 1$ includes self and nearest neighbor interactions, while $ \text{dist} = 2$ consists of self, nearest, and next-nearest neighbor terms. A schematic can be found in Fig.~\ref{SSH_fig}. Notice there is an asymmetry in the intersite interactions. That is, for $ \text{dist} = 1$ the first series in (\ref{wannier_psi_b2}) includes $a_{m}$ and $a_{m+1}$, but not $a_{m-1}$. Similarly, the second series of  (\ref{wannier_psi_a2}) contains $b_m$ and $b_{m-1}$, but not $b_{m+1}$.
In addition, we have replaced the continuous eigenvalue $\mu$ with $\lambda$ to distinguish the two. To be clear, the continuous eigenvalue $\mu$ is satisfies (\ref{Schrodinger_eqn}) and  the discrete tight-binding eigenvalue $\lambda$ satisfies (\ref{wannier_psi_a2})-(\ref{wannier_psi_b2}). The goal is to make these two quantities as close as possible. 



Let us consider  system (\ref{wannier_psi_a2})-(\ref{wannier_psi_b2}) with self and  nearest neighbor interactions (${\rm dist} = 1$); this is the simplest nontrivial tight-binding model. Using the periodic nature of the potential and its symmetries, as well as the translation properties of Wannier functions, one obtains the well-known SSH model
 \begin{equation}
\label{SSH1}
    \begin{aligned}
        fa_n + cb_n + db_{n-1} &= \lambda a_n , \\
        fb_n + ca_n + da_{n+1} &= \lambda b_n ,
    \end{aligned}
\end{equation}
for $n \in \mathbb{Z}$,  where $f $ is a self-interaction coefficient and $c, d $ are the nearest neighbor interactions (see Fig.~\ref{SSH_fig}). For a real potential $V(x)$ in (\ref{Schrodinger_eqn}), these coefficients are real. In the case of photonic waveguide arrays, the coefficient strengths are inversely related to the distance between lattice sites. Note that the on-site interaction term $f$ can be removed by shifting the spectrum, but we leave it in to ensure proper fitting. We review some of the analytical properties of the SSH model in Appendix~\ref{SSH_review_sec}.

This approach can be extended to include more terms. For example, next-nearest (dist = 2) and next-next-nearest (dist = 3) neighbor interactions retained. However, in the deep lattice limit, i.e. $V_0 \rightarrow \infty$, these nonlocal terms become exponentially small. 

The essential theme of this work is finding the optimal choice of coefficients. For instance, what is the optimal choice of coefficients $f,c,d$  in (\ref{SSH1}) to minimize the difference between spectral bands $\mu$ and $\lambda$? One approach is to numerically compute the Wannier functions and evaluate the inner products that form the coefficients (see Sec.~\ref{wannier_comparison}). This process can be expensive and tedious. The next section introduces the novel contribution of this work, a minimization approach to building tight-binding models 
that show agreement with the continuous bands {\it and} exhibit the appropriate topology.

\section{An Optimization Approach}
\label{optimize_sec}

In this section 
an optimization algorithm for determining the coefficients of tight-binding models is implemented. There are several benefits to this approach. First, this optimization approach is automated; no integrals need to be computed. Second, for a modest amount of coefficients. e.g. 3-9, and a reasonable initial guess, the algorithm often converges in a few seconds. Third,  one can extend the number of interactions to achieve models that approximate the original equation to arbitrary levels of precision, like a spectral method.

To determine the optimal  interaction coefficients, a  least squares approach is implemented.
An   objective function that consists  of the difference between the (given) continuous bands generated from the Schr{\"o}dinger equation (\ref{bloch_wave_eqn}) and the discrete band approximation is to be minimized.
Since the discrete bands (eigenvalues) are nonlinear functions of the coefficients, this is a nonlinear least squares optimization problem.
The Levenberg-Marquardt algorithm is implemented to minimize the total residual error.

This method is able to approximate the bulk spectral bands to arbitrary precision. 
Furthermore, the discrete models are imbued with the appropriate topology, inherited from the original differential equation. 
Finally, in addition to approximating the spectral bulk bands and topology, we also desire this algorithm to produce models that are able to approximate any edge modes. The performance of this optimization algorithm is examined in Section \ref{results}.

To begin, let us introduce the optimization problem. The discrete model we seek has real coefficients that act as tuning parameters. As an example, recall the SSH model in (\ref{SSH}) which has three coefficients: $f,c,d$. These are the unknowns we are solving for. Assume there are $r$ 
unknown coefficients to solve for; denote them by $x_1, x_2 , \dots x_r$ 
and store them in the vector ${\bf X} = (x_1, x_2 , \dots, x_r)^T \in \mathbb{R}^{r}$. 

Define the residual functions
\begin{equation}
\label{residual_func}
    \begin{aligned}
        r_1(k; {\bf X}) &= \lambda_1(k; {\bf X} ) - \mu_1(k) , \\
        r_2(k; {\bf X})&= \lambda_2(k; {\bf X} ) - \mu_2(k) , 
    \end{aligned}
\end{equation}
corresponding to the first and second spectral bands, respectively. 
Here, $\mu_1(k)$ and $\mu_2(k)$ correspond to given spectral bands  (see Fig.~\ref{Schrodinger_bands_plot}). These could be numerically computed or possibly experimentally generated. In this work these are obtained by numerically solving the Schr{\"o}dinger equation (\ref{bloch_wave_eqn}). The values of $\lambda_1$ and $\lambda_2$ denote the discrete approximations; they correspond to the discrete bulk spectral band approximations. A simple example of these latter functions is given in (\ref{SSH_bulk_bands_define}) for the 
SSH model.

Next, we  generate  equations to solve for the unknown coefficients. Sample the quasi-momentum values of the first Brillouin zone $k \in \left[-\tfrac{\pi}{L}, \tfrac{\pi}{L}\right]$ in the Schr\"odinger equation (\ref{bloch_wave_eqn}) at $m $ distinct points: $k_1, k_2 , \dots , k_{m}$, with $k_i \not= k_j$ for $i \not= j$. The simplest way to do this is to take a uniform spacing: $k_j = - \tfrac{\pi}{L} + (j-1)  \Delta k, j = 1, 2 ,  \dots, m $ for spacing $\Delta k = \tfrac{2 \pi}{m L}$. Store these spectral values in the vector ${\bf k} = (k_1, k_2 , \dots, k_m)^T \in \mathbb{R}^{m}$.

We wish to numerically determine  the  coefficients that  minimize  a total residual function that combines $r_1(k)$ and $r_2(k)$. 
Discretize the residual functions (\ref{residual_func}) by the following residual vectors at the sample points: $\mathbf{r}_1 \equiv r_1({\bf k}; {\bf X})$ and $\mathbf{r}_2 \equiv r_2({\bf k}; {\bf X})$, or
\begin{equation}
    \mathbf{r}_1 = \begin{pmatrix}
        r_1(k_1; {\bf X})\\
        r_1(k_2; {\bf X})\\
        \vdots\\
        r_1(k_m; {\bf X})
    \end{pmatrix} , \hspace{1em} \mathbf{r}_2 = \begin{pmatrix}
        r_2(k_1; {\bf X})\\
        r_2(k_2; {\bf X})\\
        \vdots\\
        r_2(k_m; {\bf X})
    \end{pmatrix} ,
\end{equation}
such that $\mathbf{r}_1, \mathbf{r}_2 \in \mathbb{R}^{m}$. Define the  objective functions  to be minimized, one for each band 
\begin{equation}
\label{objective_banded}
    \begin{aligned}
        F_1 = \frac{1}{2}\|\mathbf{r}_1\|^2_2 ,  ~~~~~~~~
        F_2 = \frac{1}{2}\|\mathbf{r}_2\|^2_2 ,
    \end{aligned}
\end{equation}
in terms of the $\ell^2$ norm $|| {\bf v} ||_2^2  = v_1^2 + v_2^2 + \cdots v_m^2$.

Experience shows that it is important to incorporate both bands into the total objective function and weight them equally. Optimizing only one band tends to favor the approximation of that band at the expense of the other. When applying the algorithm to each band individually, we were unable to find consistent convergence for both bands.  To obtain the best overall fit in general, we find it best to construct the super vector $\mathbf{R} \in \mathbb{R}^{2m}$ as
\begin{equation}
\label{super_residual}
    \mathbf{R} = \begin{pmatrix} \mathbf{r}_1\\ \mathbf{r}_2 \end{pmatrix} .
\end{equation}
We apply our optimization algorithm to this \say{super} residual, $\mathbf{R}$, using the total objective function
\begin{equation}
    \label{super_objective}
    F = \frac{1}{2}\|\mathbf{R}\|^2_2 = \frac{1}{2}\|\mathbf{r}_1\|^2_2 +  \frac{1}{2}\|\mathbf{r}_2\|^2_2  .
\end{equation}

Note that if we wanted to favor certain data points over others, positive weight factors could be added to certain terms of the super-residual (\ref{super_residual}). As an example, suppose one felt more confident about the measurement obtained for $\mu_2$, the second band. Then one could modify the total residual (\ref{super_objective}) by 
\begin{equation*}
    \label{super_objective_weight}
    F \rightarrow  \frac{1}{2} \left( w_1\|\mathbf{r}_1\|^2_2 +  w_2 \|\mathbf{r}_2\|^2_2 \right) ,
\end{equation*}
where $w_2 > w_1 > 0 $; this would favor the terms corresponding to the second band. All results presented below used the equally weighted objective function in (\ref{super_objective}), i.e. $w_1 = w_2 = 1$.

Next the algorithm used to minimize the super objective function (\ref{super_objective}) is discussed. A Levenberg–Marquardt (LM) algorithm is found to be an effective manner  of solving this the nonlinear least squares problem. The LM method is a well-known and established method for these types of problems \cite{Nocedal2006}.


\subsection{The Levenberg–Marquardt method} 
\label{Levenberg–Marquardt}



We start from the super-objective function (\ref{super_objective}). 
Following a formulation similar to \cite{Nocedal2006}, define the residual gradients 
\begin{align*}
    {\bm \nabla} r_1(k;{\bf X}) = \begin{pmatrix}
            \frac{\partial r_1}{\partial x_1} (k;{\bf X}) \\
            \\
            \frac{\partial r_1}{\partial x_2} (k;{\bf X}) \\
            \\
            \vdots\\
            \\
            \frac{\partial r_1}{\partial x_r} (k;{\bf X})
        \end{pmatrix}, ~~~~ {\bm \nabla} r_2(k;{\bf X}) = \begin{pmatrix}
            \frac{\partial r_2}{\partial x_1} (k;{\bf X}) \\
            \\
            \frac{\partial r_2}{\partial x_2} (k;{\bf X}) \\
            \\
            \vdots\\
            \\
            \frac{\partial r_2}{\partial x_r} (k;{\bf X})
        \end{pmatrix} , 
\end{align*}
where $ {\bm \nabla} r_{1,2} \in \mathbb{R}^{r}$.
The Jacobian matrix of each band has the following form
\begin{equation*}
\label{jacobians}
    J_1({\bf X}) = \begin{pmatrix}
        {\bm \nabla} r_1(k_1;{\bf X})^T\\
        {\bm \nabla} r_1(k_2;{\bf X})^T\\
        \vdots\\
        {\bm \nabla} r_1(k_m;{\bf X})^T
    \end{pmatrix}, \hspace{1.2em}
    J_2({\bf X}) = \begin{pmatrix}
        {\bm \nabla} r_2(k_1;{\bf X})^T\\
        {\bm \nabla} r_2(k_2;{\bf X})^T\\
        \vdots\\
        {\bm \nabla} r_2(k_m;{\bf X})^T
    \end{pmatrix},
\end{equation*}
where $J_{1,2} \in \mathbb{R}^{m \times r}$. The gradient for the objective function of each band in (\ref{objective_banded}) is
\begin{equation*}
    \begin{aligned}
        {\bm \nabla} F_1 = J_1({\bf X})^T {\bf r}_1({\bf X})  ,  ~~~~
        {\bm \nabla} F_2 = J_2({\bf X})^T {\bf r}_2({\bf X})  ,
    \end{aligned}
\end{equation*}
where ${\bm \nabla} F_{1,2} \in \mathbb{R}^r$. The corresponding Hessians are approximated by 
\begin{equation*}
    \begin{aligned}
        \nabla^2 F_1 & = J_1({\bf X})^T  J_1({\bf X})  + \sum_{j = 1}^m r_1(k_j) \nabla^2 r_1(k_j) \approx J_1({\bf X})^T  J_1({\bf X})  , \\
        \nabla^2 F_2 & = J_2({\bf X})^T  J_2({\bf X})  + \sum_{j = 1}^m r_2(k_j) \nabla^2 r_2(k_j)  \approx J_2({\bf X})^T  J_2({\bf X}) ,
    \end{aligned}
\end{equation*}
where $ \nabla^2 F_{1,2} \in \mathbb{R}^{r \times r}$. This is the Gauss-Newton  (GN) simplification, it avoids having to compute any second derivative residual terms. Neglecting the second-order derivative terms  is justified since typically either (a) the residuals $r_{1,2}$ are  small, or (b) the residuals are nearly affine and $\nabla^2 r_{1,2}$ are small.
Finally, let us combine both bands into a single method. Define the super-Jacobian
\begin{equation}
\label{super_jacobi}
    \mathbf{J} = \begin{pmatrix} J_1\\ J_2 \end{pmatrix} ,
\end{equation}
where ${\bf J} \in \mathbb{R}^{2m \times r}$.

The LM method update at the $j^{\rm th}$ iteration  is given by 
\begin{equation}
\label{LM_method}
    \mathbf{X}_{j + 1} = \mathbf{X}_j - \left[\mathbf{J}(\mathbf{X}_j)^T\mathbf{J}(\mathbf{X}_j) + \rho \mathbb{I}\right]^{-1}\mathbf{J}(\mathbf{X}_j)^T\mathbf{R}_j ,
\end{equation}
where $\rho > 0 $ is a parameter to be specified below and $\mathbb{I}$ is the identity matrix.  This method can be viewed as an interpolation between the Gauss-Newton and steepest descent algorithms. 
At each iteration, we adopt the following dynamic update of the parameter $\rho$:
\begin{itemize}
\item if $\|\mathbf{R}_{j+1}\|_2^2 \geq \|\mathbf{R}_{j}\|_2^2$, then for some fixed parameter $\nu > 1$, we update 
\[\rho \to \nu \rho.\]
\item Otherwise, if $\|\mathbf{R}_{j+1}\|_2^2 < \|\mathbf{R}_{j}\|_2^2$, then we update
\[\rho \to \frac{\rho}{\nu}.\]
\end{itemize}
More sophisticated methods for updating $\rho$ may be utilized, but we found consistent and often rapid convergence results using this simple approach. 
For our implementation, we chose $\rho_0 = 5$ (initial value of $\rho$) and $\nu = 1.1$.

In our experience, the advantage of using LM over Gauss-Newton for this project is that LM typically converged to a 
solutions in only a few (less than 5) iterations, whereas GN could take 
upwards of hundreds of 
iterations to find a solution for certain problems.

As an example, consider the three coefficient SSH model (\ref{SSH}). The unknowns are $x_1 = f, x_2 = c, x_3 = d$, or ${\bf X} = (f,c,d)$. 
The discrete bands $\lambda_{1,2}(k;f,c,d)$  given in (\ref{SSH_bulk_bands_define}) give the residual functions defined in (\ref{residual_func})
\begin{equation}
\label{SSH_residuals}
    \begin{aligned}
    r_1(k) = f + |c + de^{ik}| - \mu_1(k) , \\
     r_2(k) =   f - |c + de^{ik}| - \mu_2(k) ,
    \end{aligned}
\end{equation}
where $\mu_{1,2}(k)$ are given data, e.g. numerics in Fig.~\ref{Schrodinger_bands_plot}.  From these residuals, the super-residual in (\ref{super_residual}) can be formed by evaluating at the $m$ sample points in $k$, i.e. ${\bf r} _{1,2} = r_{1,2}({\bf k})$.
The gradients of $r_1$ and $r_2$ with respect to the unknown coefficients are:
\begin{equation*}
    \begin{aligned}
        {\bm \nabla} r_1(k;f,c,d) &= \begin{pmatrix}
            \frac{\partial r_1}{\partial f}\\
            \\
            \frac{\partial r_1}{\partial c}\\
            \\
            \frac{\partial r_1}{\partial d}
             \end{pmatrix}
            = \begin{pmatrix}
            1\\
            \\
            \tfrac{c + d\cos(k)}{|c + de^{ik}|} \\
            \\
            \tfrac{d + c\cos(k)}{|c + de^{ik}|}
        \end{pmatrix}, \\
        {\bm \nabla} r_2(k;f,c,d) &= \begin{pmatrix}
            \frac{\partial r_2}{\partial f}\\
            \\
            \frac{\partial r_2}{\partial c}\\
            \\
            \frac{\partial r_2}{\partial d}
        \end{pmatrix} 
        = \begin{pmatrix}
            1\\
            \\
            -\tfrac{c + d\cos(k)}{|c + de^{ik}|}\\
            \\
            -\tfrac{d + c\cos(k)}{|c + de^{ik}|}
        \end{pmatrix} .
    \end{aligned}
\end{equation*}
The corresponding Jacobians are given by
\begin{equation*}
    \begin{aligned}
    J_1 &= \begin{pmatrix}
        {\bm \nabla} r_1(k_1;f,c,d)^T\\
        {\bm \nabla} r_1(k_2;f,c,d)^T\\
        \vdots\\
        {\bm \nabla} r_1(k_m;f,c,d)^T
    \end{pmatrix}
    = \begin{pmatrix}
    1 &  \tfrac{c + d\cos(k_1)}{|c + de^{ik_1}|} & \tfrac{d + c\cos(k_1)}{|c + de^{ik_1}|}  \\
    1 &   \tfrac{c + d\cos(k_2)}{|c + de^{ik_2}|}  &  \tfrac{d + c\cos(k_2)}{|c + de^{ik_2}|}  \\
     \vdots & \vdots  & \vdots \\
    1 &  \tfrac{c + d\cos(k_m)}{|c + de^{ik_m}|} &  \tfrac{d + c\cos(k_m)}{|c + de^{ik_m}|} 
    \end{pmatrix},  \\\\
    J_2 &= \begin{pmatrix}
        {\bm \nabla} r_2(k_1;f,c,d)^T\\
        {\bm \nabla} r_2(k_2;f,c,d)^T\\
        \vdots\\
        {\bm \nabla} r_2(k_m;f,c,d)^T
    \end{pmatrix}
    = \begin{pmatrix}
    1 &  - \tfrac{c + d\cos(k_1)}{|c + de^{ik_1}|} & - \tfrac{d + c\cos(k_1)}{|c + de^{ik_1}|}  \\
    1 &   - \tfrac{c + d\cos(k_2)}{|c + de^{ik_2}|}  & - \tfrac{d + c\cos(k_2)}{|c + de^{ik_2}|}  \\
     \vdots & \vdots  & \vdots \\
    1 & -  \tfrac{c + d\cos(k_m)}{|c + de^{ik_m}|} & - \tfrac{d + c\cos(k_m)}{|c + de^{ik_m}|} 
    \end{pmatrix} . 
    \end{aligned}
\end{equation*}    
From these Jacobian matrices, the super-Jacobian in (\ref{super_jacobi}) can be formed.



\subsection{Arbitrary number of coefficients}
\label{arb_coeff_sec}

Next let us discuss the form of the optimization algorithm and computations for arbitrary number of coefficients. 
First, we want to build a two-band model consisting of two identical molecules per unit cell. We wish to preserve the underlying symmetries of both the Schr\"odinger (\ref{Schrodinger_eqn}) and classic SSH  (\ref{SSH}) models. Namely, we wish for this model to preserve the inversion and chiral (up to a spectral shift) symmetries. 

 \subsubsection{Extended  bulk problem}

\begin{figure}
    \includegraphics[scale = 0.3]{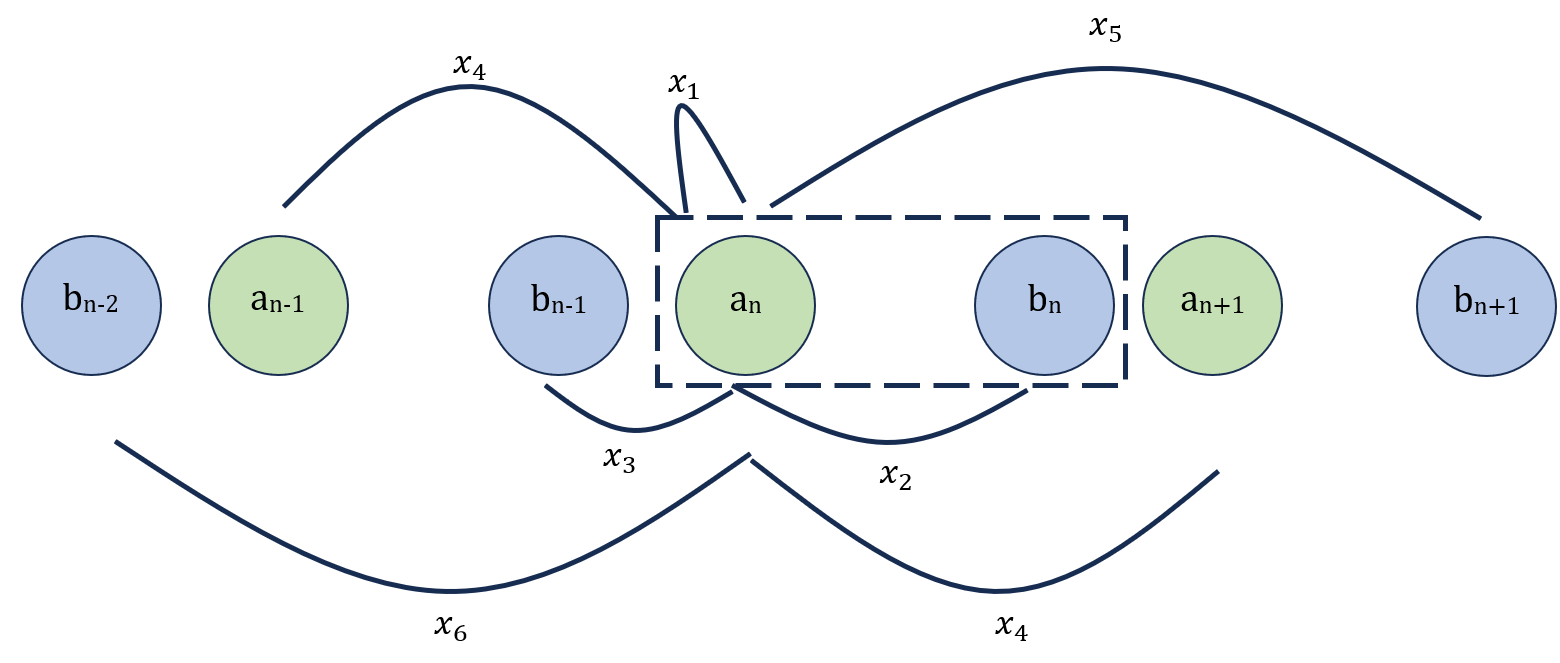}
    \centering
    \caption{Example of the extended SSH model with 6 interaction coefficients. The pattern continues for multiples of three coefficients.  \label{extended_SSH}}
\end{figure}

To simplify the process of imposing symmetries, we 
restrict our attention to  multiplies of three coefficients. This means each update takes an interaction of the same type (intramolecular), e.g. $a$-site with another $a$-site, and an intermolecular interaction between $a$-sites and $b$-sites. The three coefficient model was examined in (\ref{SSH}) and Fig.~\ref{SSH_fig}.  The general extended SSH equations, which include infinitely many interactions, were derived in (\ref{wannier_psi_a2})-(\ref{wannier_psi_b2}) and are given by
    \begin{align}
        \label{general_extended_SSH_model}
\sum_{j = -\infty}^{\infty} x_{3 |j| + 1} a_{n + j} + \sum_{j = 0}^{\infty} x_{3 j + 2} b_{n + j} + \sum_{j = 1}^{\infty}  x_{3 j} b_{n - j} & =  \lambda a_n , \nonumber \\ 
\sum_{j = 0}^{\infty} x_{3 j + 2} a_{n - j} + \sum_{j = 1}^{\infty}  x_{3 j} a_{n + j} + \sum_{j = -\infty}^{\infty} x_{3 |j| + 1} b_{n + j} & =  \lambda b_n ,
    \end{align}
where $x_j \in \mathbb{R}$ are the model coefficients and  $|j|$ denotes absolute value of $j$. 
Taking these interactions preserves the  inversion symmetry 
of the model. A diagram depicting a six coefficient model is shown in Fig.~\ref{extended_SSH}. In a six coefficient model, for example, the $a_n$ site interacts with $a_n, b_{n}, b_{n-1},a_{n \pm 1}, b_{n+1}, b_{n - 2}$. Note that the next-nearest neighbor $x_4$ is the same for both $a_{n \pm 1}$.
Explicitly, the governing equations are given by 
\begin{eqnarray}
&& x_1 a_n + x_2 b_n + x_3 b_{n-1} + x_4 \left( a_{n+1} + a_{n-1} \right) + x_5 b_{n+1}+ x_6 b_{n-2}   \nonumber\\
&&=\lambda a_n  , \nonumber\\
&& x_1 b_n + x_2 a_n + x_3 a_{n+1} + x_4 \left( b_{n+1} + b_{n-1} \right) + x_5 a_{n-1} + x_6 a_{n+2}  \nonumber\\
&&=\lambda b_n .
\end{eqnarray}

For the remainder of this section, there will be $r = 3 \kappa$ interactions, where $\kappa$ is some positive integer ($\kappa = 1$ is the classic SSH model in Appendix~\ref{SSH_review_sec}, $\kappa = 2$ in the six interaction case described above). 
The coefficients $x_{3j + 1} , j = 0, 1, \dots , \kappa - 1$ denote intramolecular interactions, i.e. $a$ with $a$ or $b$ with $b$. The coefficients $x_{3j + 2}, j = 0, 1, \dots , \kappa - 1 $ denote interactions to the right (left) for $a$ $(b)$ sites. The coefficients $x_{3j}, j = 1, 2, \dots, \kappa$ denote interactions to the left (right) for $a$ $(b)$ sites.
Since these coefficients are directly related to distance, each of the three sequences: $x_1, x_4 , x_ 7 , \dots $, and $x_2, x_5 , x_ 8, \dots$, and $x_3, x_6 , x_ 9 , \dots$ should descend in magnitude, that is, $x_{3j}, x_{3j + 1} , x_{3j + 2} \rightarrow 0$ as $j \rightarrow \infty$. This is not surprising and is the root of the tight-binding approximation.  Any potential solution that violates this descending 
nature is considered non-physical since it suggesting nonlocal sites are unreasonably large. In general, it is difficult to conclude relative magnitudes among the different sequences.

Next, we look for plane wave solutions of the form  (\ref{SSH_plane_waves}). The generalized bulk SSH system  is given by 
\begin{widetext}
    \begin{equation}
   \label{general_discrete_bulk_system}
    \begin{pmatrix}
        {\displaystyle x_1 + \sum_{j=1}^{\kappa-1}2x_{3j+1}\cos(jkL)} & {\displaystyle \sum_{j=1}^{\kappa}\left(x_{3j-1}e^{i(j-1)kL} + x_{3j}e^{-ijkL}\right)}\\\\
        {\displaystyle \sum_{j=1}^{\kappa}\left(x_{3j-1} {e^{-i(j - 1)kL}} + x_{3j}e^{ijkL}\right)} & {\displaystyle x_1 + \sum_{j=1}^{\kappa-1}2x_{3j+1}\cos(jkL)}
    \end{pmatrix}\begin{pmatrix}
        \alpha\\
        \beta
    \end{pmatrix} = \lambda(k)\begin{pmatrix}
        \alpha\\
        \beta
    \end{pmatrix} ,
    \end{equation}
\end{widetext}
for $\kappa = 1,2,3, \dots (r = 3,6,9, \dots)$.
All self-interactions are symmetric, that is the coefficients at $a_{n + p}$ and $a_{n - p}$ are the same for any $p \in \mathbb{Z}^+$; this symmetry is similar for the $b$-sites. As a result, the diagonal terms involve cosines whose arguments are proportional to the  displacements. The associated eigenvalues (bands) are given by
\begin{eqnarray}
\label{general_discrete_band}
    \lambda_{1,2}(k) &&= \left(x_1 + \sum_{j=1}^{\kappa-1}2x_{3j+1}\cos(jkL)\right) \nonumber\\
    &&\pm \left|\sum_{j=1}^{\kappa}\left(x_{3j-1}e^{i(j-1)kL} + x_{3j}e^{-ijkL}\right)\right| ,
\end{eqnarray}
where $\lambda_1 (\lambda_2)$ corresponds to the upper (lower) branch.
These are the discrete approximations used in the residual functions (\ref{residual_func}).

As an example, consider the  six-coefficient configuration shown in Fig.~\ref{extended_SSH}. Here, the bulk Hamiltonian ($r = 6$) is given by
\begin{align*}
{H} & = \begin{pmatrix} {H}_{11} & {H}_{12} \\ {H}_{12}^* & {H}_{11}    \end{pmatrix} , \\
{H}_{11}  & = x_1 + 2x_{4} \cos(kL)  , \\
{H}_{12}  & = x_2 + x_3 e^{- i kL } + x_5 e^{ i k L} + x_6 e^{- 2i k L} ,  \\
\end{align*}
and the corresponding spectral bands (eigenvalues) are 
\begin{widetext}
    \begin{align*}
    \label{6_coeff_discrete_band}
        \lambda_{1,2}(k)  = x_1 + 2x_{4}\cos(kL) 
      \pm   \sqrt{x_2^2 + x_3^2 + x_5^2 + x_6^2 + 2 \left( x_2 x_3 + x_2 x_5 + x_3 x_6 \right) \cos (k L ) + 2 \left(  x_2 x_6 + x_3 x_5\right) \cos (2 k L) + 2 x_5 x_6 \cos (3 k L )}.
    \end{align*}
\end{widetext}

Computing the derivatives of (\ref{general_discrete_band}) to form the Jacobian matrices in (\ref{jacobians}) is a tedious task. Some authors approximate these terms numerically using a finite difference approximation \cite{Nocedal2006}. However, this immediately introduces an error, often first or second order.
Here, we compute the Jacobian matrices symbolically using MATLAB's built in symbolic differentiation toolbox. This is an essential step for ensuring a user-friendly interaction. Moreover, this means the elements of the Jacobian matrices are {\it exact}. On the other hand, when the number of coefficients approaches 20, this is the most time-consuming portion of this method. We have found that for higher numbers of coefficients (21 or more), the total run time for the algorithm can take 30 or more minutes. 


\subsubsection{Extended edge problem}
\label{extended_ssh_edge_sec}

Next we discuss the computation of edge modes in the extended SSH model.
Localized edge modes exist when a nontrivial topology is present in the bulk spectral eigenfunctions, this is the so-called bulk-edge correspondence. The bulk (infinite) problem is truncated to a finite (but large) domain by imposing the open/Dirichlet zero boundary conditions:
\begin{equation*}
a_n, b_n = 0 , ~~~~ n < 1~~ \text{and}~~ n > N ,
\end{equation*}
where $N \gg 1$. Typically, we set $N = 100$. This is the setup for computing edge modes, localized modes at the endpoints that decay exponentially away from the boundary. That is, at the left endpoint an edge mode satisfies $a_n, b_n \rightarrow 0$ as $n \rightarrow \infty$; a similar mode exists on the right endpoint.

Physically, the presence of edge modes corresponds to defect lattice sites at an endpoint. Everywhere interior the lattice sites appear in dimer pairs, a coupled set of lattice sites. On the other hand, the topologically trivial case corresponds to a complete dimerization where all potential sites come in pairs.

The block matrix system for arbitrary coefficients is formed by creating $N \times N$ 
Toeplitz matrices with the interaction terms on the diagonals. Each row represents the equations (\ref{general_extended_SSH_model}) evaluated at a different value of $n$. Self-interactions occur on the diagonal while super and sub-diagonals become more and more nonlocal as they move from the main diagonal. The total number of nonzero diagonals (upper and lower) is $r$, 
the number of interaction coefficients.

Define the vector
\begin{equation}
{\bf v} = (a_1, a_2 , \dots, a_N | b_1, b_2, \dots, b_N)^T ,
\end{equation}
that contains all nontrivial solution values in the interval $n \in \{1 ,  2 , \dots, N \}$.
The self-interaction terms (both a-sites and b-sites) are on alternating diagonal elements of $\mathcal{H}_{11}$
\begin{equation*}
    \mathcal{H}_{11} = \begin{pmatrix}
        x_1 & x_4 & \cdots & x_{3\kappa - 2} & & & \\
        x_4 & x_1 & x_4 & \ddots & \ddots & & \\
        \vdots & x_4 & x_1 & \ddots & \ddots & \ddots & \\
        x_{3\kappa - 2} & \ddots & \ddots & \ddots &\ddots & \ddots & x_{3\kappa - 2}\\
        & \ddots & \ddots & \ddots & \ddots & \ddots & \vdots\\
        & & \ddots & \ddots & \ddots & \ddots & x_4\\
        & & & x_{3\kappa - 2} & \cdots & x_4 & x_1
    \end{pmatrix} .
\end{equation*}
Next, the intersite interactions (b-sites effect on a-sites) are placed on alternating diagonals of $\mathcal{H}_{12}$ and have the following  form
\begin{equation*}
    \mathcal{H}_{12} = \begin{pmatrix}
        x_2 & x_5 & \cdots & x_{3\kappa -1} & & & \\
        x_3 & x_2 & x_5 & \ddots & \ddots & & \\
        \vdots & x_3 & x_2 & \ddots & \ddots & \ddots &  \\
        x_{3\kappa} & \ddots & \ddots & \ddots & \ddots & \ddots & x_{3\kappa-1} \\
        & \ddots & \ddots & \ddots & \ddots & \ddots & \vdots\\
        & & \ddots & \ddots & \ddots & \ddots & x_5\\
        & & & x_{3\kappa} & \cdots & x_3 & x_2
    \end{pmatrix} .
\end{equation*}
Due to the inherent inversion symmetry of the SSH model, the interaction of a-sites with b-sites is the transpose of this matrix.
Thus, the edge problem is solved by computing the eigenmodes of the eigenvalue problem
\begin{equation}
\label{edge_extended_prob}
 \mathcal{H} {\bf v} = \lambda {\bf v} , ~~~~ 
    \mathcal{H} = \begin{bmatrix}
        \mathcal{H}_{11} & \mathcal{H}_{12}\\
        \mathcal{H}_{12}^T & \mathcal{H}_{11}
    \end{bmatrix} ,
\end{equation}
for the $2N \times 2N$ 
block matrix $\mathcal{H}$. Note that the matrix $\mathcal{H}$ is Hermitian and so $\lambda \in \mathbb{R}$.
Upon 
solution of (\ref{edge_extended_prob}), the sites are re-ordered as: $\cdots , a_{n-1}, b_{n-1}, a_n , b_n, a_{n+1}, b_{n+1} , \cdots$.

\subsubsection{Iteration considerations}

A well-chosen initial guess is critical for convergence to a solution with the correct topology. 
The importance can be traced back to Fig.~\ref{Schrodinger_bands_plot} where two topologically distinct lattice configurations yield identical spectral bands. This also occurs in the classical SSH model, that is, the bands in the left and right panels of Fig.~\ref{SSH_bands_plot} are identical. The spectral band data is not enough to uniquely determine the topology of the system. Some additional piece of data, e.g. eigenfunctions, is needed. 

To ensure we converge to solutions with appropriate topology, we consider initial guesses of the form
\begin{equation}
\label{IC_LM}
 {\bf X}_0 = \left( f  , c, d , 0  , \dots , 0 \right)^T ,
\end{equation}
where $f,c,d \in \mathbb{R}$. Depending on the desired topology, we adjust the relative magnitude of the coefficients. Specifically, for  a nontopological model (null Zak phase), we choose an initial guess with $c > d > 0$. For a  a topological model (nonzero Zak phase), we choose initial guess with $0 < c < d$. The value of $f$ is chosen so that $f > c$ and $f > d$. Intuitively this makes sense because this self-interaction term is the most local (and strongest) of all interactions. Choosing the initial guesses this way, we find this always converges to a solution with the appropriate topology; there is no need to impose topology on iterative method beyond this initial guess.

The exiting criterion taken is when the relative two-norm  difference between successive iterations is lower than some specified tolerance, $\varepsilon$, that is
\begin{equation}
\label{kickout_criterion}
    \frac{ \|{\bf X}_{j+1} - {\bf X}_{j}\|_2}{\| {\bf X}_{j+1} \|_2} < \varepsilon .
\end{equation}
The tolerance is left as an input for the user. The reason for this is that 
the tolerance must be adjusted 
to ensure sufficient convergence of the algorithm. For instance, choosing a relatively large tolerance (e.g. $\varepsilon = 10^{-3}$) for one system may be sufficient to find a local minimum (that is, decreasing the tolerance will not result in a sizable decrease in the residual), but the same tolerance for higher numbers of coefficients or a larger potential depth may cause the algorithm to stop before it finds a suitable local minimum. Moreover, for relatively large tolerances, the algorithm may 
converge to a non-physical solution, whereas for smaller tolerances, the algorithms converges to a physically reasonable solution. A few examples are detailed in the next section. 

When 
 condition (\ref{kickout_criterion}) is met, the algorithm stops. A typical starting tolerance value is $\varepsilon  = 10^{-3}$. Even with this rather large tolerance, we are able to find objective (residual) values of (\ref{super_objective}) on the order of $10^{-8}$ for three and six coefficients. These values are typical for deep potentials ($V_0^2 \approx 150$), but more shallow potentials (i.e. $V_0^2 \approx 50$) a smaller tolerance may be required. 

A predetermined maximum number of iterations is hard-coded into the program. This value is set to 400. 
The LM method typically converges in 10 iterations or less.  
While the LM method typically converges in a few iterations, the maximum number of iterations is set as large as it is to cover cases where a significant number of iterations may be required for the program to terminate. This may be the case when decreasing the tolerance to very small values (e.g. $\varepsilon \approx 10^{-16}$).

\section{Results}
\label{results}
In this section the results of our optimization approach  are presented. Overall, we find that the program is able to generate  discrete models that describe  one dimensional SSH-type topological insulator systems. More specifically, these models are capable of  reproducing the expected spectral and topological properties. Further, we find that increasing the potential depth or increasing the number of interaction coefficients increases the accuracy of the fit to arbitrary accuracy. Due to the automatic nature of our method, increasing the number of coefficients is a trivial request.

For each computation presented in this section, we fix the number of spectral points at $m = 64$. That is, each band contains $m$ equally spaced sample points in the Brillouin zone. The tolerance we select depends on the value of the total residual; we discuss the tolerance we choose in more detail below.  As the total residual decreases, the tolerance $\varepsilon$ used in (\ref{kickout_criterion}) needs to decrease. In practice we typically take a tolerance of $\varepsilon = 10^{-3}$ for potential depths around $V_0^2 \approx 150$ 
for three and six coefficients. When increasing the number of coefficients above six (9,12,...), we decrease the tolerance by roughly an order of magnitude for each increase in the number of coefficients. That is, we may take $\varepsilon = 10^{-4}$ for 9 coefficients and $\varepsilon = 10^{-5}$ for 12 coefficients. Of course, choosing a tolerance is going to depend on the given system and the number of coefficients chosen.

\subsection{A first example}
Here we inspect the output 
of the algorithm when given the typical SSH bands (\ref{SSH_bulk_bands_define}) with 
prescribed coefficient values,  $f,c,d$. That is, we consider the performance of the method for a cooked example. Since the algorithm constructs the SSH bands explicitly, we 
 expect 
 the algorithm should rapidly find the prescribed coefficients. We inspect the performance of the algorithm when decreasing the tolerance. As a note, since we are providing the SSH bands, we must tell the algorithm what the associated topology is.

\begin{figure}
        \hspace{0.5cm}\includegraphics[scale = 0.18]{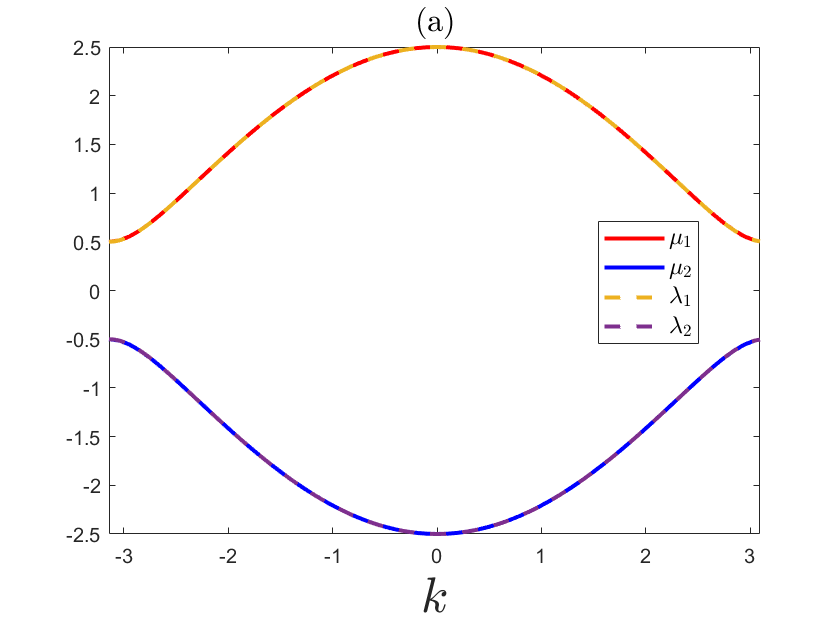}
        \includegraphics[scale = 0.18]{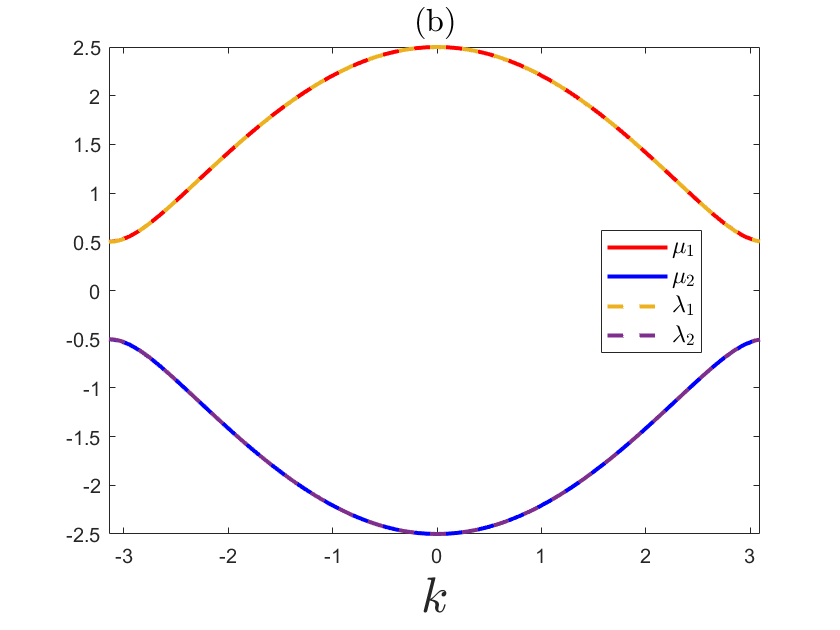}
        \newline
        \includegraphics[scale = 0.18]{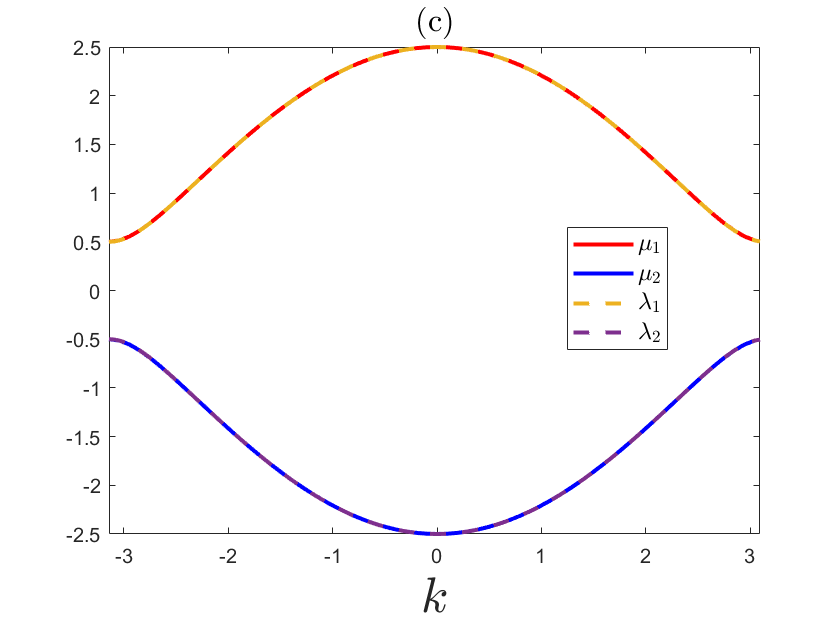}
    \centering
    \caption{Comparison between the exact, $\mu_{1,2}$, and approximate, $\lambda_{1,2}$, spectral bands. Top row: Three coefficient band fit for prescribed coefficients $f = 0$, $c = 1$, $d = 1.5$  given in (\ref{SSH_bulk_bands_define}) with (a) $\varepsilon = 10^{-3}$ and (b) $\varepsilon = 10^{-6}$. (c) 6 coefficient fit with the same prescribed coefficients as in (a) and (b) with $\varepsilon = 10^{-6}$  
    \label{prescribed_fit_eps_3}}
\end{figure}

 For this example, we take $f = 0$, $c = 1$, $d = 1.5$, which corresponds to a topologically nontrivial state (see (\ref{zak_calc_SSH})). A comparison between the exact and approximate bands is shown in Fig.~\ref{prescribed_fit_eps_3}; 
 the difference between the bands is imperceptible to the naked eye.
Running the program with a tolerance of $\varepsilon = 10^{-3}$, we find the coefficients shown in Table~\ref{table_cooked_ex_3coeff} with associated residual of $ \frac{1}{2} || {\bf R} ||_2^2 \approx 7.19 \times 10^{-5}$. Examining the data, we see that each absolute difference is on the order of $\mathcal{O}(10^{-5})$ to $\mathcal{O}(10^{-4})$. More generally, the accuracy of the coefficients appears to be at  or near the tolerance level. Said differently, the algorithm only works as hard as it has to.  This is highlighted in Table~\ref{table_cooked_ex_3coeff_lower_tol} when we decrease the relative tolerance $\varepsilon = 10^{-6}$. Here the total residual decreases to  $\frac{1}{2} || {\bf R} ||_2^2 \approx 5.78 \times 10^{-11}$. The absolute difference is on the order of $\mathcal{O}(10^{-7})$ or better. Lastly, these results clearly have the correct topology, i.e. nontrivial. 
\begin{table}
\centering
    \begin{tabular}{||c|c|c||}
        \hline
        Prescribed Coeffs & Optimization Coeffs & Absolute Diff\\
        \hline\hline
        $x_1, f = 0$ & 1.996782$\times 10^{-5}$ & 1.996782$\times 10^{-5}$\\
        \hline
        $x_2, c = 1$ & 0.9985484476 & 0.00014515524\\
        \hline
        $x_3, d = 1.5$ & 1.50010877876 & 0.00010877876\\
        \hline
    \end{tabular}
\caption{Outputs for the spectral band inputs  given in (\ref{SSH_bulk_bands_define}) with values $f = 0, c = 1, d = 1.5$  for relative tolerance $\varepsilon = 10^{-3}$. \label{table_cooked_ex_3coeff}}
\end{table}

The previous example teaches us that: (i) this method is quite effective at finding the appropriate coefficients and  (ii) the accuracy of these coefficients is directly related to the relative tolerance cutoff used to terminate the iterative method given in (\ref{kickout_criterion}).

\begin{table}
\centering
    \begin{tabular}{||c|c|c||}
        \hline
         Prescribed Coeffs & Optimization Coeffs & Absolute Diff\\
         \hline\hline
         $x_1, f = 0$ & 4.62680$\times 10^{-10}$ & 4.62680$\times 10^{-10}$\\
         \hline
         $x_2, c = 1$ & 0.99999986868 & 1.31323$\times 10^{-7}$\\
         \hline
         $x_3, d = 1.5$ & 1.50000008783 & 8.78337$\times 10^{-8}$\\
         \hline
    \end{tabular}
\caption{Outputs for the spectral band inputs  given in (\ref{SSH_bulk_bands_define}) with values $f = 0, c = 1, d = 1.5$ for relative tolerance $\varepsilon = 10^{-6}$. \label{table_cooked_ex_3coeff_lower_tol}}
\end{table}

Next, we consider how the method performs when the number of fitting coefficients increases. In this case, we continue examining the classic SSH model bands given in (\ref{SSH_bulk_bands_define}) with six coefficients. To be clear, this approximation should only have two nontrivial coefficients, i.e. $x_2,x_3 $, and we  expect the remaining  coefficients should be close to zero. The discrete approximation of the spectral bands is compared with the test case in Fig.~\ref{prescribed_fit_eps_3}; the curves are indistinguishable to the naked eye. A comparison of the fitted coefficient values is given in Table~\ref{table_cooked_ex_6coeff}. As one can see, the difference between the prescribed and fitted coefficients is order  $\mathcal{O}(10^{-5})$ or better. Again, these values are near the tolerance level.  In this case, when a coefficient is expected to be null, the algorithm does find it (to within the tolerance level).


\begin{table}
\centering
    \begin{tabular}{||c|c|c||}
        \hline
        Prescribed Coeffs & Optimization Coeffs & Absolute Diff\\
        \hline\hline
        $x_1 = 0$ & 0 & 0\\
        \hline
        $x_2 = 1$ & 0.99997901271 & 2.09873$\times 10^{-5}$\\
        \hline
        $x_3 = 1.5$ & 1.50001399110 & 1.39911$\times 10^{-5}$\\
        \hline
        $x_4 = 0$ & 0 & 0\\
        \hline
        $x_5 = 0$ & -1.3991102587$\times 10^{-5}$ & 1.39911$\times 10^{-5}$\\
        \hline
        $x_6 = 0$ & 2.0987093293$\times 10^{-5}$ & 2.09871$\times 10^{-5}$\\
        \hline
    \end{tabular}
\caption{Outputs for the spectral band inputs  given in (\ref{SSH_bulk_bands_define}) with values $f = 0, c = 1, d = 1.5$ for relative tolerance $\varepsilon = 10^{-3}$. \label{table_cooked_ex_6coeff}}
\end{table}

\subsubsection{Increasing potential depth (fixed interactions)}
\begin{figure}
    \includegraphics[scale = 0.19]{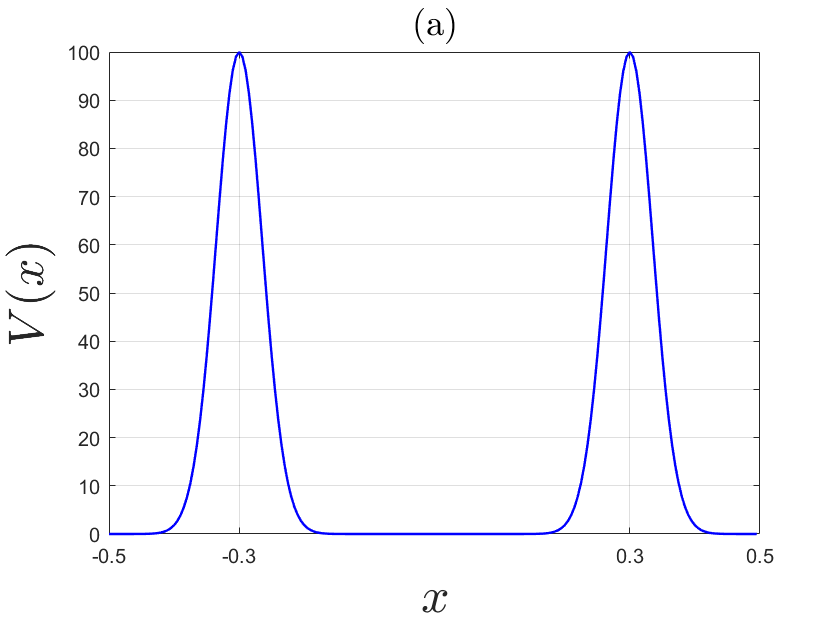}
    \includegraphics[scale = 0.19]{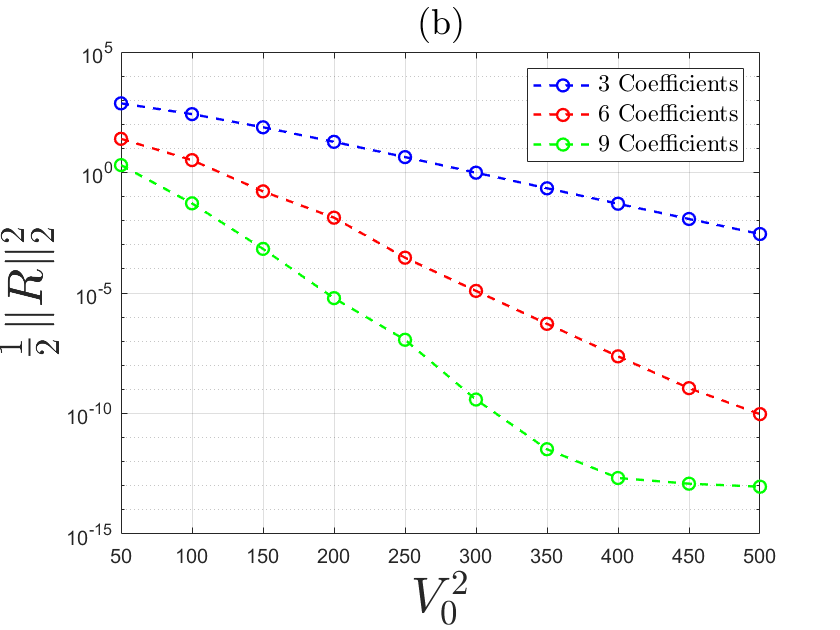}
    \centering
    \caption{(a) One period (L = 1) of periodic potential (\ref{pot_expand}) with parameters: $\sigma = 0.05,  x_1 = -0.3$, $x_2 = 0.3$, and $V_0^2 = 100$.
    (b) Total residual (\ref{super_objective}) for three, six, and nine coefficients ($r = 3,6,9$) varying the potential depth. All parameters are the same except $50 \leq V_0^2 \leq 500$. \label{resid_vs_pot}}
\end{figure}

\begin{figure}
    \includegraphics[scale = 0.19]{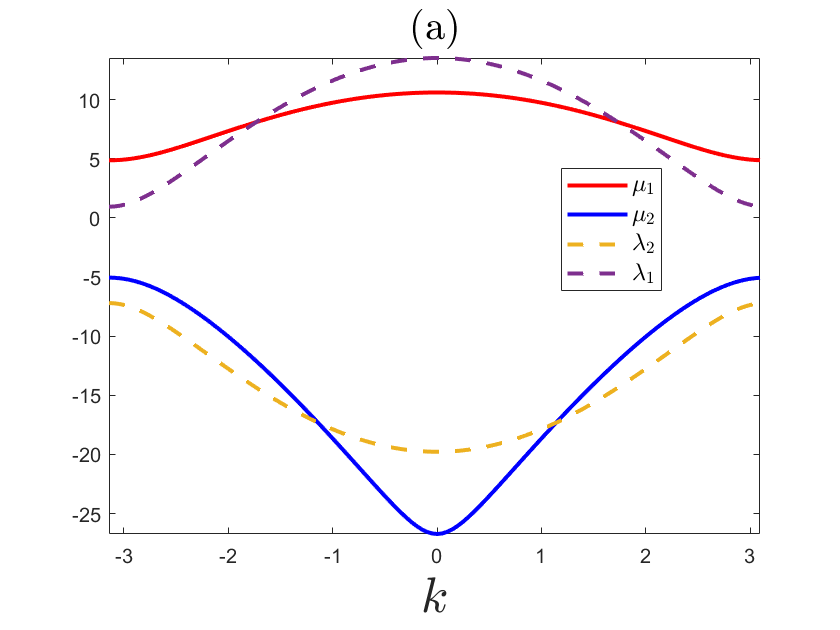}
    \includegraphics[scale = 0.19]{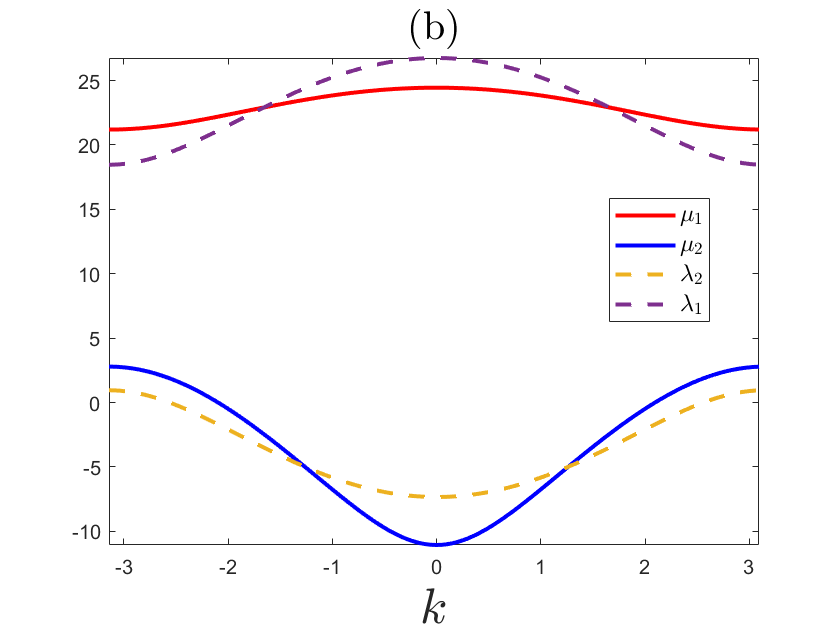}
    \centering
    \\
    \includegraphics[scale = 0.19]{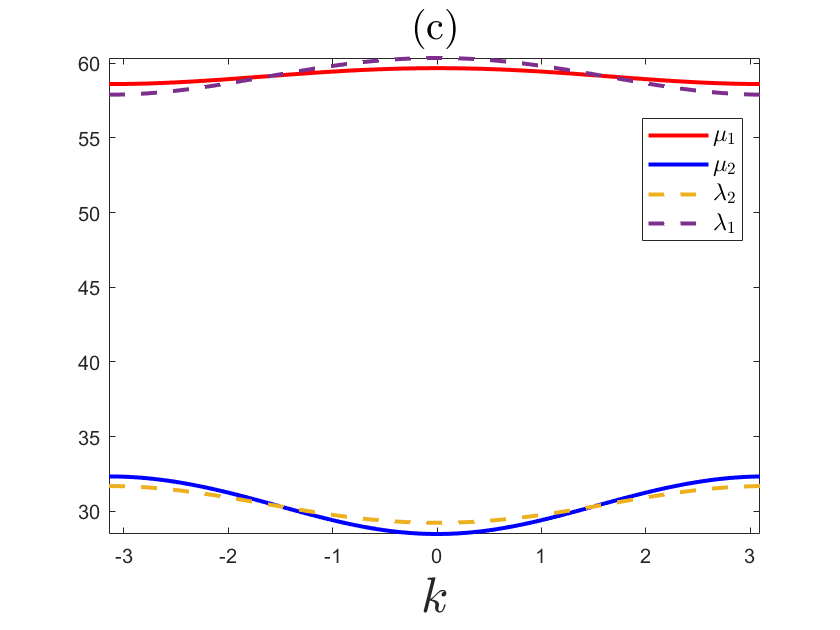}
    \includegraphics[scale = 0.19]{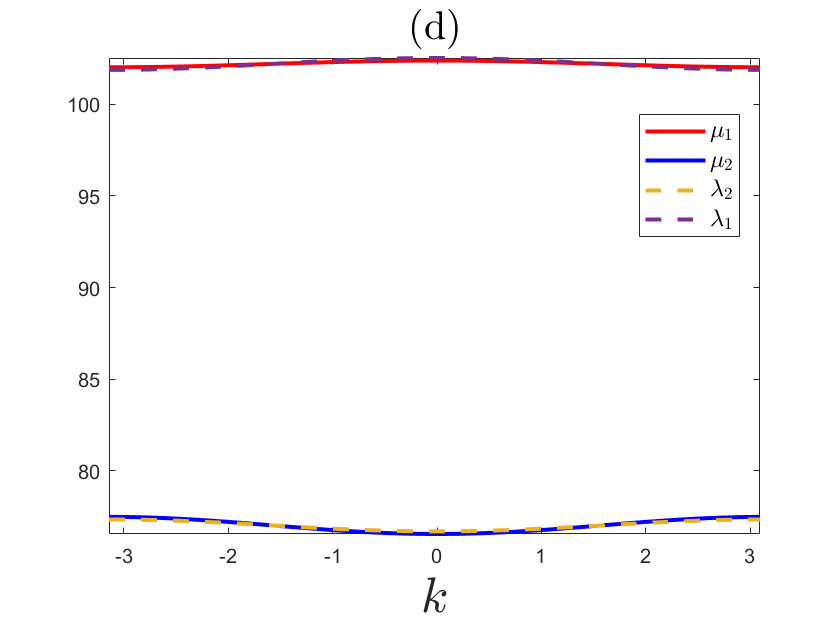}
    \centering
    \caption{Band fits for three coefficients ($x_j = 0, j > 3$) and increasing potential depths.  The potential depths are: (a)  $V_0^2 = 50$, (b)  $V_0^2 = 100$, (c)  $V_0^2 = 200$, (d) $V_0^2 = 300$. All other parameters  are the same as those  in Fig.~\ref{resid_vs_pot}.
     \label{3coeff_fits}}
\end{figure}

Next, we consider fits as the lattice depth increases. That is, consider  potential (\ref{pot_expand}) as $V_0^2 \rightarrow \infty$. One period ($L = 1$) of the potential is shown in Fig.~\ref{resid_vs_pot}(a).  A couple typical cases of the spectral band fits are shown in Fig.~\ref{3coeff_fits} as the lattice depth increases. Notice that as the potential depth increases, the bands tend to flatten out. Each case considered here uses three coefficients ($r = 3$), i.e. the ``classic SSH model'' considered in Appendix~\ref{SSH_review_sec}, and by $V_0^2 = 300$ the difference between the continuous and discrete bands is difficult to distinguish. In these figures and below, ``continuous'' refers to the bands obtained by solving the Schr\"odinger equation (\ref{bloch_wave_eqn}), while ``discrete'' refers to the tight-binding approximation given in (\ref{general_discrete_band}).

In the deep lattice limit, the tight-binding approximation can rigorously be shown to converge to the continuous solution of the Schr\"odinger equation \cite{Fefferman2018,Ablowitz2012}. Intuitively, this makes sense because as the lattice depth increases, the potential lattice traps the wavefunction and the Wannier or orbital tails decay more rapidly. Specifically, in the deep lattice limit, the spectral bands flatten out, resulting in good fits with only a few coefficients.  We observe convergence, i.e. $ ||{\bf R}|| \rightarrow 0$, as the potential depth increases, i.e. $V_0^2 \rightarrow \infty$. Note the exponential decrease in the residual as $V_0$ increases in Fig.~\ref{resid_vs_pot}(b).  For $V_0^2 = 500$, we observe residual errors of $\mathcal{O}(10^{-3}), \mathcal{O}(10^{-10}), \mathcal{O}(10^{-13})$ for $r = 3, 6, 9$ coefficients, respectively.


\begin{figure}
    \includegraphics[scale = 0.19]{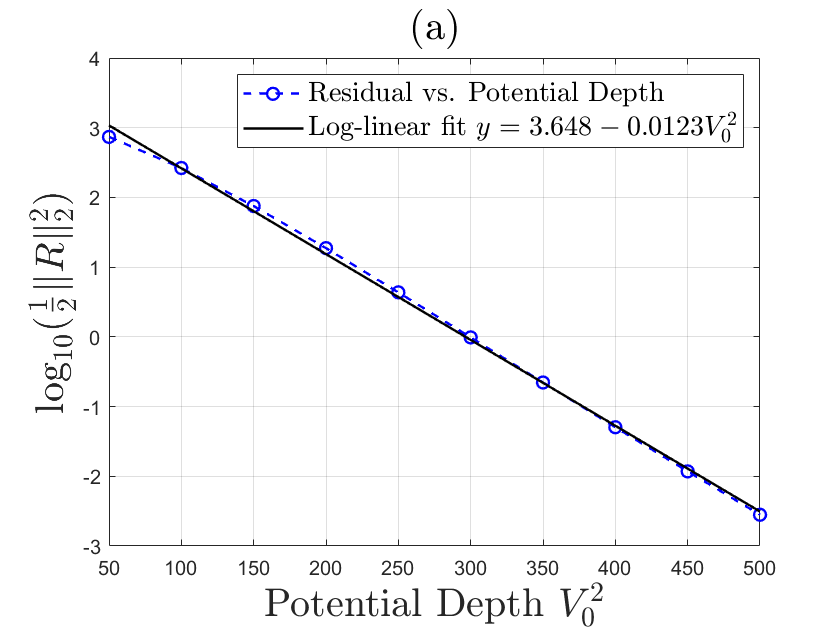}
    \includegraphics[scale = 0.19]{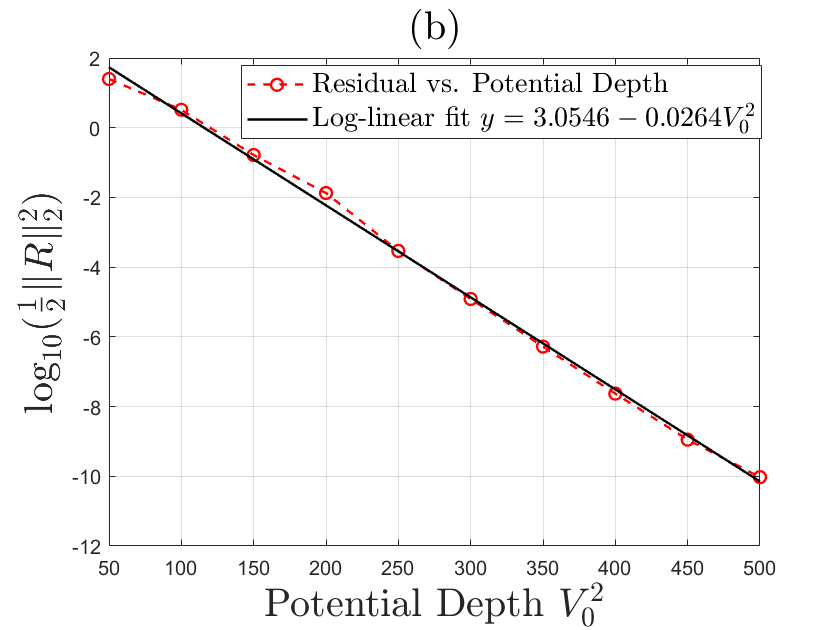}
    \includegraphics[scale = 0.19]{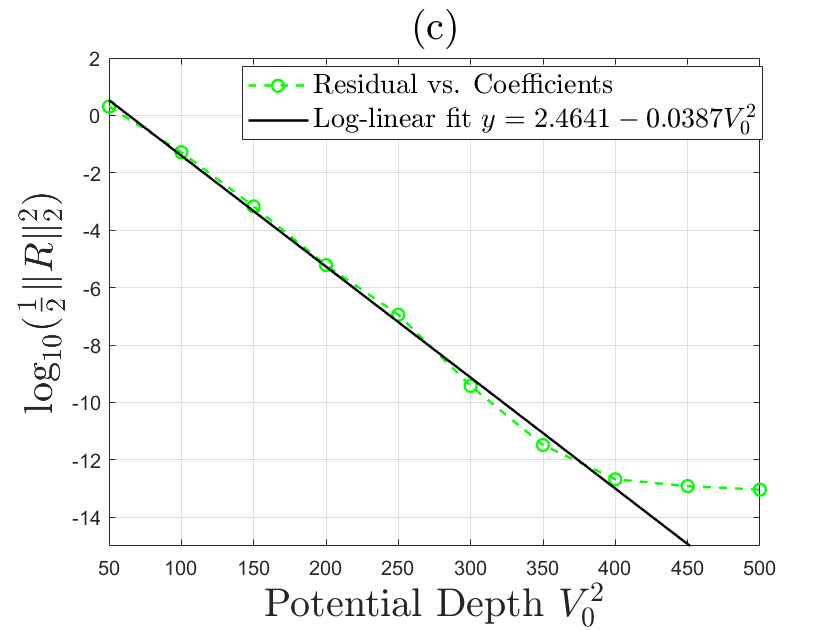}
    \centering
    \caption{Log-linear fits for residuals of (a) 3, (b) 6, and (c) 9 coefficients versus the potential depth. The tolerances changed as the potential increased and the number of coefficients increased. These are the same curves shown in Fig.~\ref{resid_vs_pot}(b). \label{converge_pot_fit}} 
\end{figure}

The convergence rate is further investigated in Fig.~\ref{converge_pot_fit} where we have applied fits to the semi-logarithmic plots. Each of these plots indicate that the total residual depends on the potential depth in the following way
\begin{equation}
\frac{1}{2} || {\bf R}||_2^2 = C t^{- V_0^2},
\end{equation}
where $C$ is a constant and $t > 1$ is the exponential base. When we take the logarithm (base 10) of both sides, we get
\begin{equation}
\log_{10} \frac{1}{2} || {\bf R}||_2^2 =\log_{10}  C - V_0^2 \log_{10}  t . 
\end{equation}
From the linear least squares fits in Fig.~(\ref{converge_pot_fit}), we infer the 
data summarized in Table~\ref{table_fit_pot_vary}.
\begin{table}[h]
\centering
\begin{tabular}{||c c c c c||} 
 \hline
$r$ &  $\log_{10} C$   & $C$ & $\log_{10} t$ & $t$ \\ [0.5ex] 
 \hline\hline
 3 & 3.648 & 4.45 $\times 10^{3}$ & 0.0123 & 1.03 \\ 
 \hline
 6 & 3.0546 & 1.13 $\times 10^{3}$ & 0.0264 & 1.06 \\
 \hline
 9 & 2.4641 & 2.91 $\times 10^2$ & 0.0387 & 1.09 \\
 \hline
\end{tabular} 
\caption{Tabular values extracted from fits in Fig.~\ref{converge_pot_fit}. \label{table_fit_pot_vary}}
\end{table}
By increasing the number of interactions, the value of $C$ is found to decrease. This is expected since including nine coefficients should give better approximations than only three, for example.
Notably, we observe the convergence rate {\it increases}, i.e. $t$ increases, as the number of coefficients increases. This says that even in the deep lattice limit, there can be value to incorporating more interactions; they can produce  faster convergence rates and reach lower errors at shallower lattice potentials.

\subsubsection{Increasing interactions (fixed potential)}

Next, consider increasing the interactions for a fixed potential configuration. Typically,  in tight-binding approximations only nearest or next-nearest neighbor interactions are included. Due to the automatic nature of our algorithm, including an arbitrary number of (nonlocal) interactions is a simple request. Additional time is required to evaluate these computations with most of the time spent computing and evaluating the Jacobian matrix.

\begin{figure}
    \includegraphics[scale = 0.19]{potential_for_inc_coeff.png}
    \includegraphics[scale = 0.19]{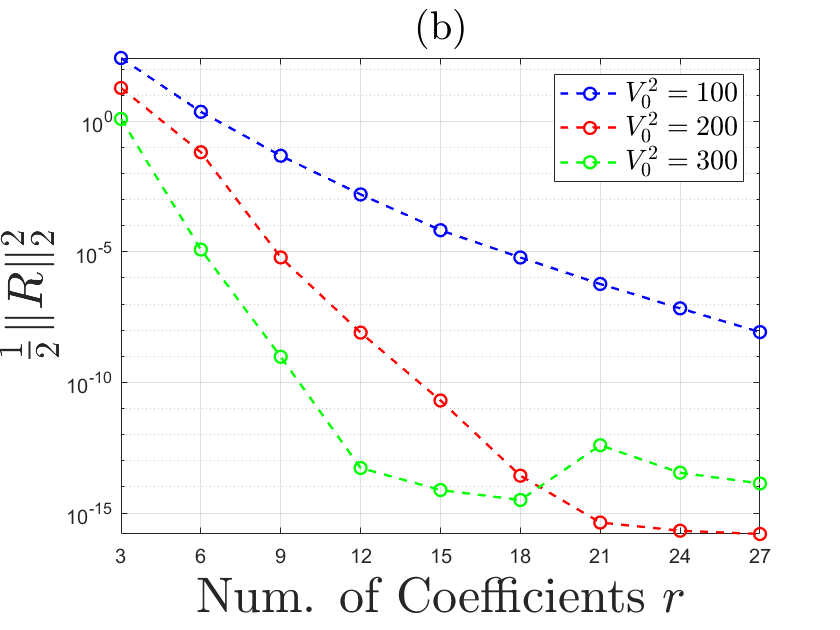}
    \centering
    \caption{(a) One period (L = 1) of periodic potential (\ref{pot_expand}) with parameters: $\sigma = 0.05,  x_1 = -0.3$, $x_2 = 0.3$, and $V_0^2 = 100$. 
    (b) Total residual (\ref{super_objective}) for $V_0^2 = 100,200,300$ as the number of interactions is increased.  \label{converge_coeff_increase}}
\end{figure}

\begin{figure}
    \includegraphics[scale = 0.19]{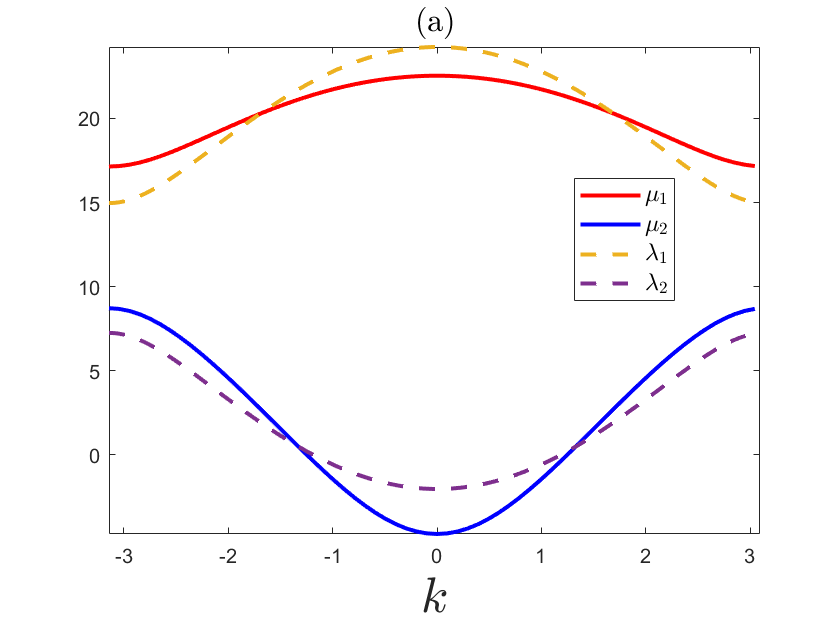}
    \includegraphics[scale = 0.19]{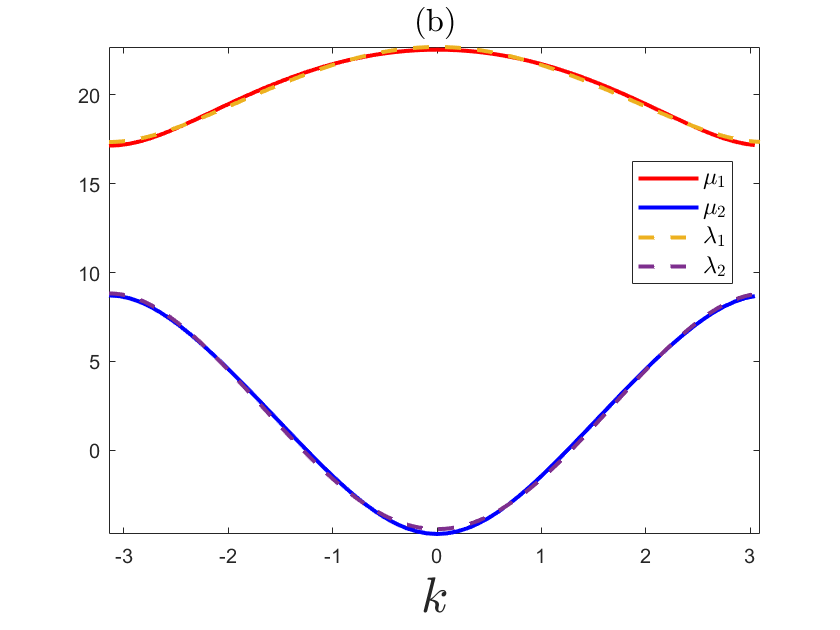}
    \newline
    \includegraphics[scale = 0.19]{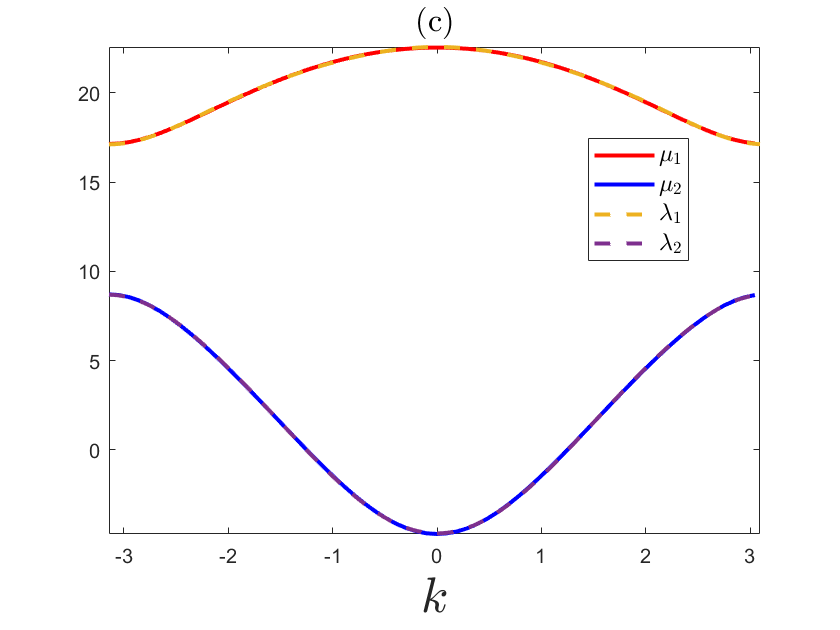}
    \includegraphics[scale = 0.19]{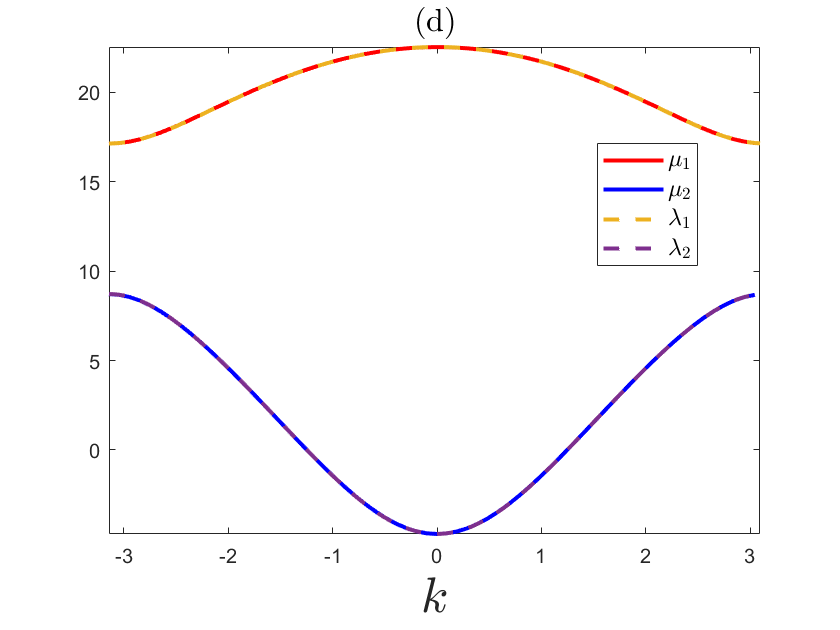}
    \centering
    \caption{Band fits as  the number of coefficients  is increased for (fixed) potential shown in Fig.~\ref{converge_coeff_increase}(a). Continuous (solid lines) and discrete (dashed lines) spectral bands for (a) $r = 3$, (b) $r = 6$, (c) $r = 9$, and (d) $r =12$ coefficients.   \label{band_fits_inc_coeff}}
\end{figure}

Consider the typical potential given in Fig.~\ref{converge_coeff_increase}(a). A couple of typical fits are shown in Fig.~\ref{band_fits_inc_coeff}. As the number of interactions increases, the discrete approximations of the bands clearly improves. For 12 coefficients the difference between the continuous and discrete bands is indistinguishable to the naked eye in a linear-linear plot. This is expected; increasing the number of interactions should improve the approximation.

Next, we explore how accurate these approximation can  become as more interactions are included. The residual error as a function of the number of interaction coefficients is shown in Fig.~\ref{converge_coeff_increase}(b) for three different potential depths. As expected, increasing the number of interactions improves the approximation.
Notice that increasing the number of coefficients yields an exponential decrease in the residual, with it bottoming out around the level of double precision round-off error, i.e. $\mathcal{O}(10^{-16})$.  

A smaller tolerance may be required when increasing the number of coefficients. For instance, running 3-6 coefficients with a tolerance of $\varepsilon = 10^{-3}$ will converge to a minimum in only a few iterations (3-5) and decreasing the tolerance will result in small improvements in the residual. Running 9 coefficients will need a smaller tolerance, e.g. $10^{-4}$. Running larger numbers of coefficients (21+) will require much smaller residuals, on the order of $\varepsilon = 10^{-8}$ to $10^{-10}$.  If the tolerance is not lowered, the method tends to converge too early and the residual error (and corresponding solution)  may not yet be optimal. 

The curves in Fig.~\ref{converge_coeff_increase}(b) reveal a dependence on the potential depth. We observe that fits for deeper potentials may also require adjusting the tolerance to find a suitable local minimum, though the relationship to potential depth does not appear to be as strong as the relationship to the number of coefficients. 
That is, for potential depths around $V_0^2 \approx 50 - 200$, the tolerances listed above are sufficient. Increasing the potential depth to around $V_0^2 \approx 300$  requires reducing the tolerance by an order of magnitude for each of the above cases. 

\begin{figure}
    \includegraphics[scale = 0.2]{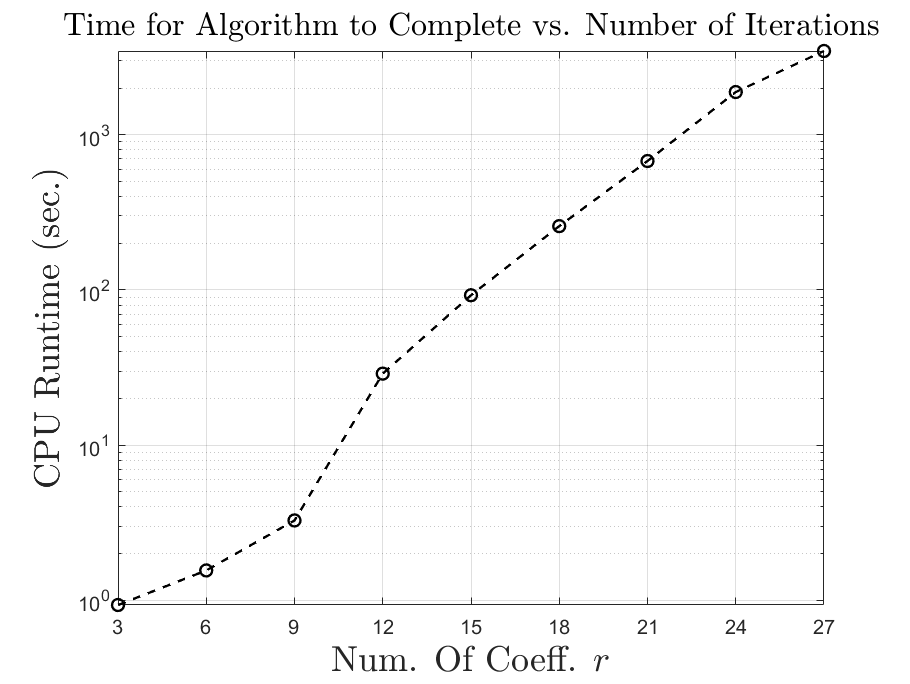}
    \centering
    \caption{CPU  runtime versus the number of coefficients. For each run, the tolerance was set to $\varepsilon = 10^{-2}$ and the algorithm converged in 
    four iterations. Note the exponential growth in the time. In each case a potential with parameters $V_0^2 = 150$, $\sigma = 0.05$, $x_1 = -0.35$, $x_2 = 0.35$ was used. \label{program_run_time}}
\end{figure}

One consequence of lowering the tolerance is the method does require more iterations and time to converge. Some typical runtime totals are shown in Fig.~\ref{program_run_time}.
These CPU times include all aspects of this optimization algorithm, including computing the Jacobian (\ref{super_jacobi}) and updating the LM method (\ref{LM_method}), until convergence is reached. For few coefficients, e.g. 3 or 6, the method often converges in a few seconds with a tolerance of $\varepsilon = 10^{-3}$. At the opposite end of the spectrum, including 24 or 27 interaction terms can take upwards of 20-30 minutes on a standard laptop with the following specifications: 12th Gen Intel i5-1240P 1.70 GHz 12 core/16 logical processor CPU, 32 GB DDR4-3200 MHz RAM. Note that in these latter cases we decrease the tolerance  (e.g. $\varepsilon = 10^{-10}$) to ensure a suitable residual value is obtained. 

\begin{figure}
    \includegraphics[scale = 0.19]{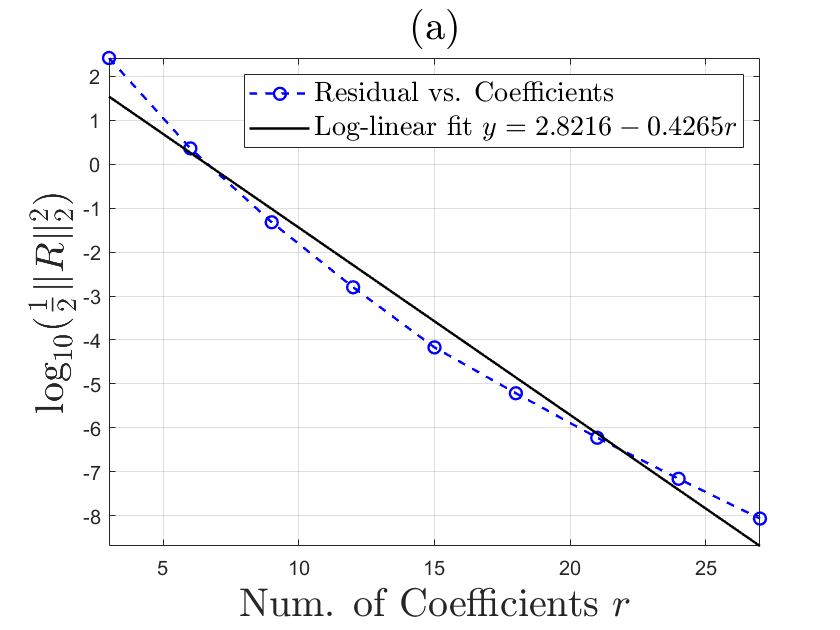}
    \includegraphics[scale = 0.19]{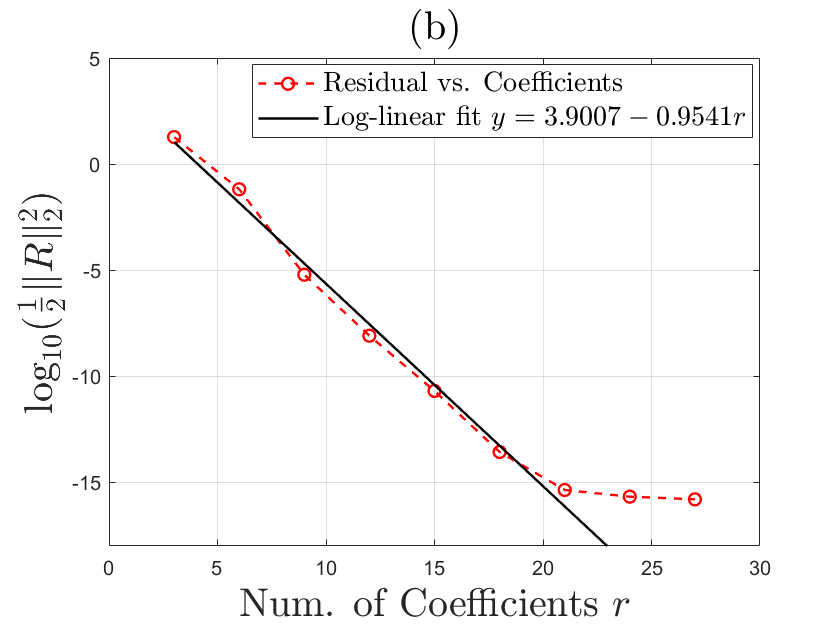}
    \includegraphics[scale = 0.19]{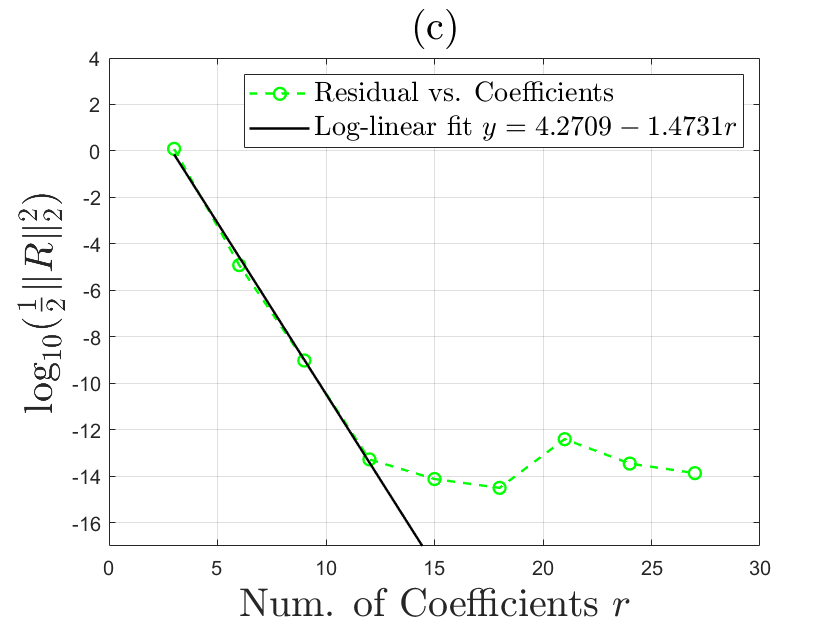}
    \centering
    \caption{Log-linear fits shown in Fig.~\ref{converge_coeff_increase}(b) for residuals of (a) $V_0^2 = 100$, (b) $V_0^2 = 200$, (c) and $V_0^2 = 300$ versus the number of coefficients.  The tolerance for these figures was also reduced as the potential depth increased. These are the same curves shown in Fig.~\ref{converge_coeff_increase}(b). \label{converge_fits_coeff_vary}}
\end{figure}

We further investigate the convergence rates in Fig.~\ref{converge_fits_coeff_vary}.  These figures indicate that the  residual, at fixed $V_0^2$, depends on the number of interaction coefficients $r$ as
\begin{equation}
\frac{1}{2} || {\bf R}||_2^2 = K t^{- r},
\end{equation}
where $K$ is a constant and $t > 1$ is the exponential base. Taking the logarithm (base 10) of both sides yields
\begin{equation}
\log_{10} \frac{1}{2} || {\bf R}||_2^2 =\log_{10}  K - r \log_{10}  t . 
\end{equation}
From the linear least squares fits shown in Fig.~\ref{converge_fits_coeff_vary}, we extract  the following values summarized in Table~\ref{table_converge_coeff_fit}.
\begin{table}[h]
\centering
\begin{tabular}{||c c c c c||} 
 \hline
$V_0^2$ &  $\log_{10} K$   & $K$ & $\log_{10} t$ & $t$ \\ [0.5ex] 
 \hline\hline
 100 & 2.8216 & 6.63 $\times 10^{2}$  & 0.4265 & 2.67  \\ 
 \hline
 200 & 3.9007 & 7.96 $\times 10^{3}$ & 0.9541  & 9.00 \\
 \hline
 300 & 4.2709 & 1.87 $\times 10^{4}$  &  1.4731 & 29.72  \\
 \hline
\end{tabular}
\caption{Tabular values extracted from fits in Fig.~\ref{converge_fits_coeff_vary}. \label{table_converge_coeff_fit}}
\end{table}

The convergence results reveal a spectral  convergence rate of the method. That is, the error converges exponentially fast with number of coefficients. This is important because it means relatively few interactions are needed to achieve high accuracy. This is reminiscent of other spectral methods which are often able to achieve levels of accuracy not possible with finite difference methods \cite{Trefethen2000}. The other significant observation is that the exponential rate of convergence  increases with $V_0^2$. Notice at $V_0^2 = 300$ a relatively large base of $t \approx 30 $. In this case, each increase of three coefficients decreases the error by roughly 4-5 orders of magnitude. Hence, only 12 coefficients are needed to reach machine precision level of error, which then limits any further convergence.

\subsubsection{Topology}

A feature we expect of these models is the appropriate topology. Namely, given some input topological value, the Zak phase of this discrete model (\ref{Zak_phase_cont}) should reproduce it. In the SSH model, a fixed set spectral bands are not enough to uniquely determine the topology. For example, consider the spectral bands shown in Fig.~\ref{Schrodinger_bands_plot} obtained by solving the Schr\"odinger equation (\ref{bloch_wave_eqn}). These bands are identical even though their corresponding potentials and topologies are distinct. Often one gets information about topology from the eigenfunctions, but here we do not wish to incorporate them directly. Note that this non-uniqueness also occurs in the SSH model where the spectral bands (\ref{SSH_bulk_bands_define}) are indistinguishable, yet topologically distinct when swapping the values $(c,d) \rightarrow (d,c)$ (cf. Figs.~\ref{SSH_bands_plot}(a) and (c)).

 The desired topology is an input in our method. The algorithm prompts the user for the expected topology: 1 for nontrivial and 0 for trivial. Recall nontrivial (trivial) topology corresponds to the presence (absence) of localized boundary states. When an equation is chosen, e.g. Schr\"odinger (\ref{Schrodinger_eqn}), the (continuous) eigenfunctions can be used to determine the topology. However, this information could come from a variety of places. For example, one could perform direct numerical simulations; one could use direct numerical simulations to infer whether energy will conduct along the boundary. Alternatively, one could   take experimental measurements. That is, one could empirically measure whether the system appears to be conducting along the boundary of a finite waveguide array. Regardless of approach, this data about the Zak phase is provided to the algorithm.

Based on the provided topology, the initial condition configuration described in (\ref{IC_LM}) was found to reliably converge to models with the desired topology. That is, providing an initial guess with a trivial (nontrivial) configuration was found to consistently converge to a solution with trivial (nontrivial) topology. No other information is incorporated regarding topology.


\begin{figure}
\centering
    \includegraphics[scale = 0.19]{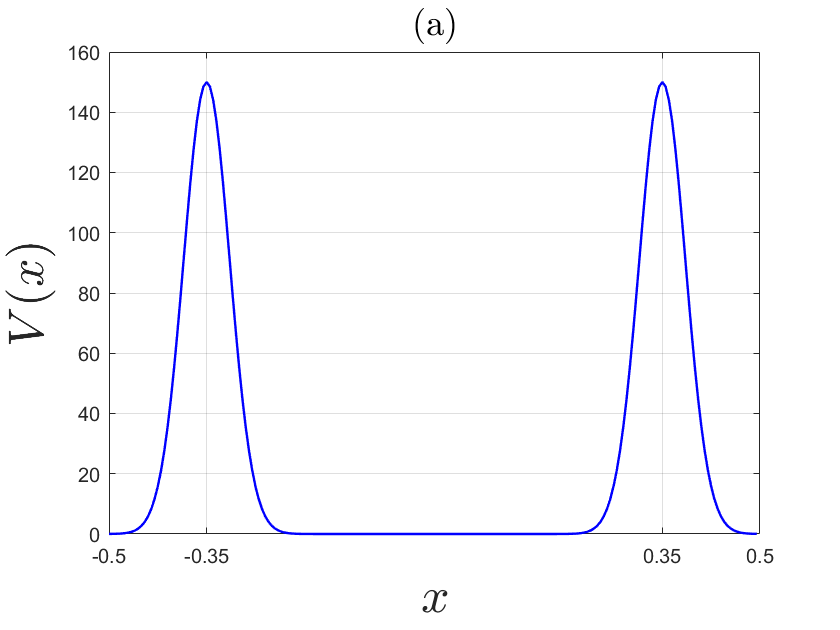}
    \includegraphics[scale = 0.19]{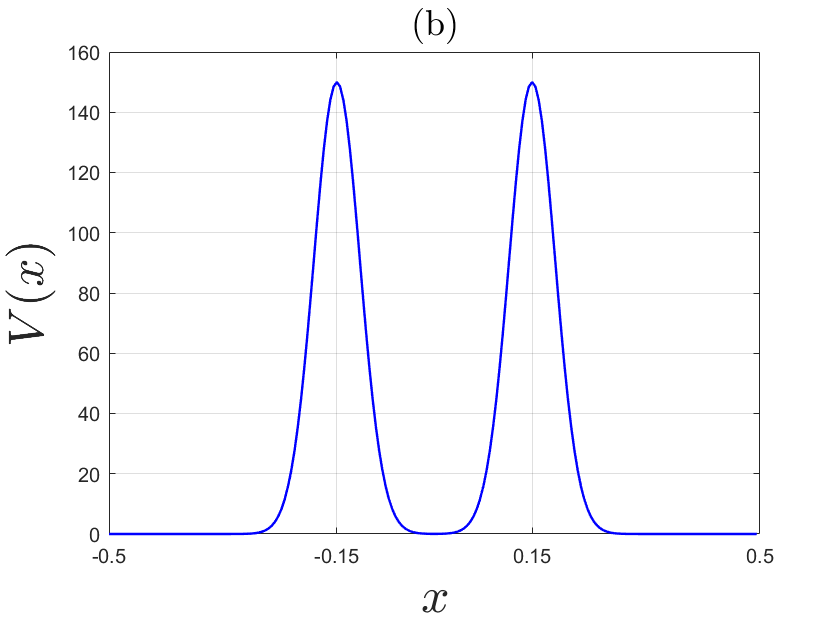} \\
    \includegraphics[scale = 0.19]{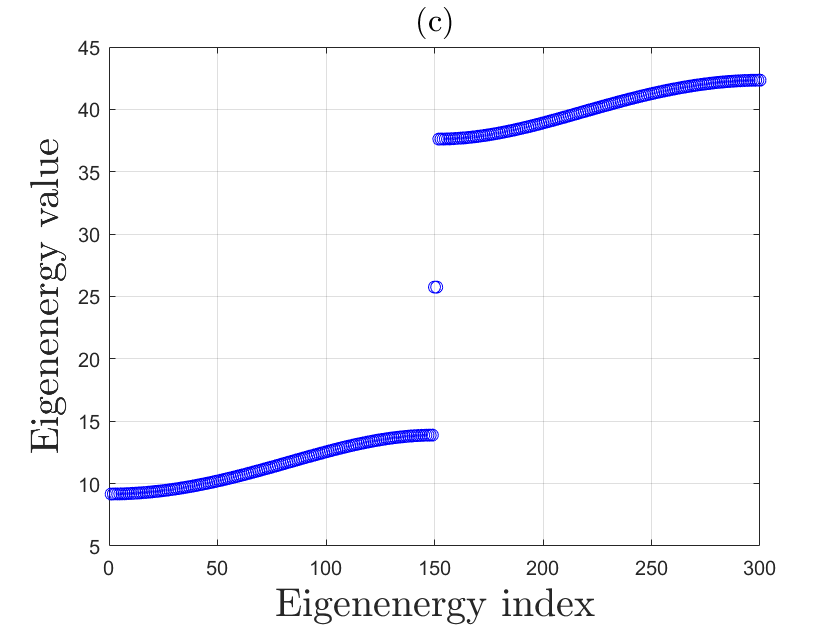}
    \includegraphics[scale = 0.19]{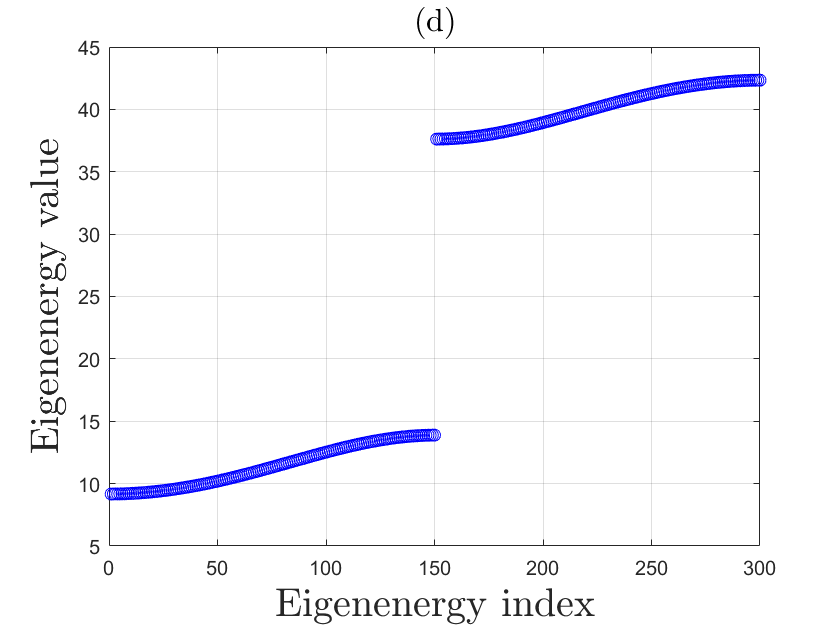}
    \caption{Top row: Potentials used in Schr\"odinger equation (\ref{bloch_wave_eqn}) and (\ref{pot_expand}) with parameters:  $V_0^2 = 150$, $\sigma = 0.05$, $L=1$, (a) $x_1 = -0.35$, $x_2 = 0.35$, and (b) $x_1 = -0.15$, $x_2 = 0.15$. Note: the corresponding spectral bands are shown Fig.~\ref{Schrodinger_bands_plot}. Bottom row: Corresponding edge spectrum obtained from (\ref{edge_extended_prob}): (c) topological (presence of edge modes) and (d) non-topological (absence of edge modes). The numerically computed Zak phases (\ref{linear_approx_zak}) are: (c) $\mathcal{Z} = \pi$ and (d) $\mathcal{Z} = 0$. \label{Discrete_band_pot_compare}}
\end{figure}

As an example, consider two topologically distinct potential configurations in Fig.~\ref{Discrete_band_pot_compare}. The potential shown in panels \ref{Discrete_band_pot_compare}(a) and \ref{Discrete_band_pot_compare}(b) correspond to nontrivial and trivial profiles, respectively. Using the continuous bands generated by these potentials (see Fig.~\ref{Schrodinger_bands_plot}), the optimization algorithm with three coefficients ($r = 3$) is applied. The corresponding edge spectrum for finite problems ($N = 100$ sites) is shown. The topological (non-topological) configurations are approximated by fitted SSH models that yield topological quantities $ \mathcal{Z} = \pi$  ($\mathcal{Z} = 0$). The topology of the discrete model is numerically computed (see (\ref{linear_approx_zak})) and is an output. Lastly, we note that the algorithm is able to consistently converge to models with the correct topology for higher numbers of coefficients (beyond nearest neighbors).

\subsubsection{Edge Modes}

Lastly, we examine the edge modes generated by our algorithm. The edge problem setup was discussed in Sec.~\ref{extended_ssh_edge_sec}. Edge modes localized on the left and right boundaries are found to coincide with nontrivial topologies. On the other hand, we do not observe any genuine edge modes when the system has a trivial topology. To identify localized edges out of the many eigenmodes, we sort the eigenvalues and look for the eigenmodes corresponding to midgap states. That is, we search for eigenmodes whose spectral frequency lies in the band gap.

\begin{figure}
    \centering
    \includegraphics[scale = 0.19]{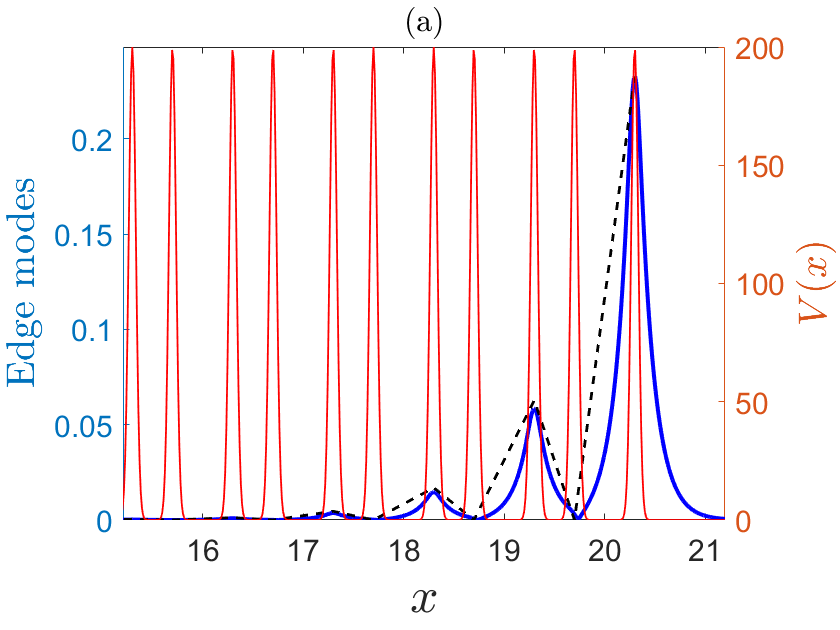}
    \includegraphics[scale = 0.19]{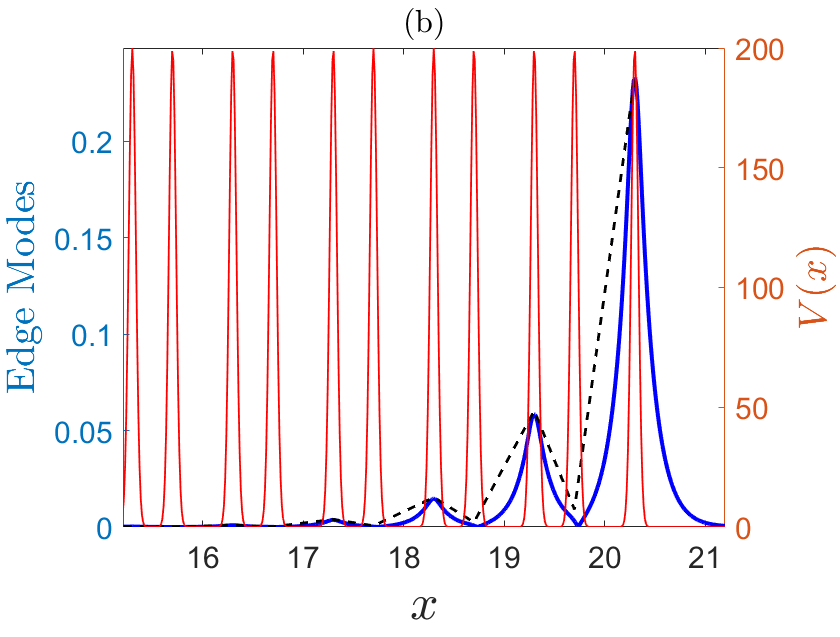}
    \newline
    \includegraphics[scale = 0.19]{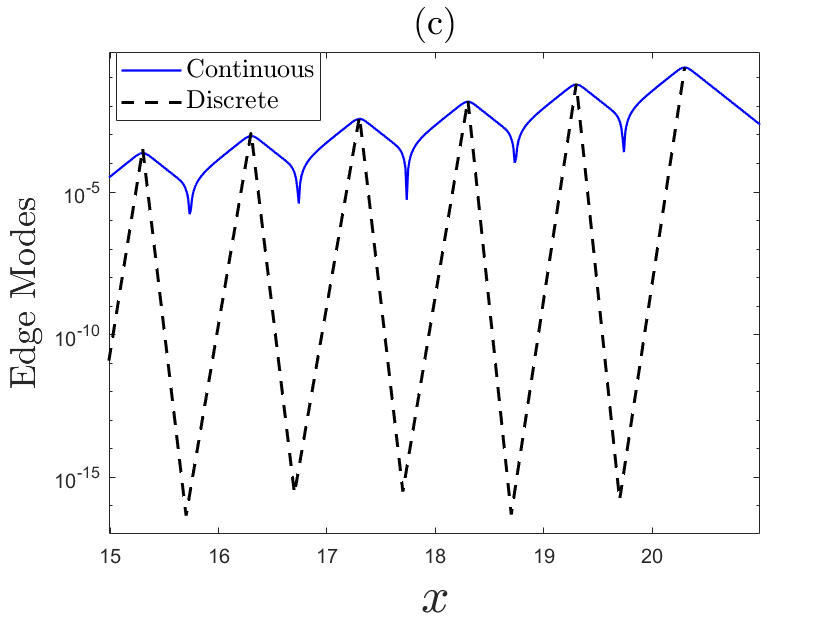}
    \includegraphics[scale = 0.19]{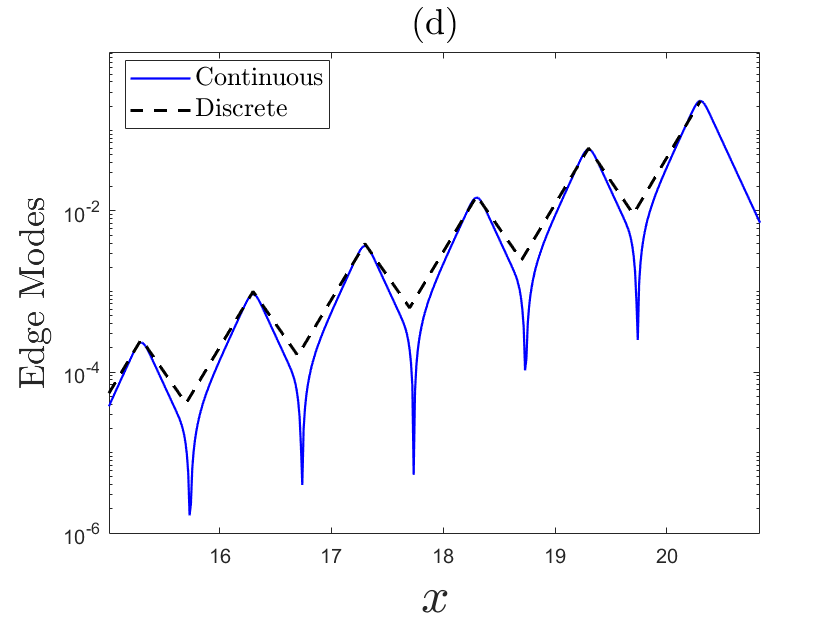}
    \caption{Comparison of continuous and discrete edge mode magnitudes plotted against the potential using (a) 3 coefficients and (b) 6 Coefficients.
    Bottom Row: Continuous and discrete edge modes plotted semi-logarithmically, where (c) corresponds to (a) and (d) corresponds to (b). The same potential was used to generate the continuous and discrete modes with the following parameters: $V_0^2 = 200$, $\sigma = 0.05$, and the peaks located at $x_1 = -0.3, ~x_2 = 0.3$. \label{cont_disc_edge_modes}}
\end{figure}

The edge modes obtained from continuous (Schr\"odinger) and discrete (extended SSH) problems are compared in Fig.~\ref{cont_disc_edge_modes}. To obtain the continuous modes, we solve (\ref{Schrodinger_eqn}) on a large computational domain and use a truncated version of the  potential (\ref{pot_expand}); the right edge of the truncated potential can be seen in Figs.~\ref{cont_disc_edge_modes}(a) and (b).
The magnitude of the edge modes is plotted on top of the potential (top row of Fig.~\ref{cont_disc_edge_modes}) to highlight their dependence on the spatial profile of the potential. Our algorithm carries with it no knowledge of the spatial (in $x$) locations of the discrete edge modes. To align these two functions, we first re-scale the (linear) discrete mode so that the largest magnitude aligns with the peak of the continuous edge mode. Next, we position the maximum and minimum values of our method to coincide with those of the continuous function. We find that the maxima occurs precisely at the right 
potential sites (i.e. $b-$sites), while the minima occur consistently occur  near the a-sites.

In the case of a three coefficient (classical) SSH model (left column of Fig.~\ref{cont_disc_edge_modes}), the peaks and overall decay of the two functions agree well. However, the continuous edge mode does not go to zero (on our spatial grid) while the discrete mode does. This 
appears exaggerated in a linear-log plot (see Fig.~\ref{cont_disc_edge_modes}(c)). Increasing the number of coefficients to six, the agreement between the two edge states improves. The improvement on the semi-logarithmic plot is quite improved (see Fig.~\ref{cont_disc_edge_modes}(d)); the maxima, decay rate, and now the minima agreement is notable.

Since no information about the edges (other than they should exist or not) is incorporated into the algorithm, this agreement is a pleasant surprise. However, in the next section we compare our coefficients to those of a standard Wannier functions expansion and note the close agreement. We conjecture this is why our discrete edge modes line up so well with the continuous edge states here.

\section{Comparison with MLWF Approach}
\label{wannier_comparison}
In Section (\ref{Wannier_Galerkin}) we derived the SSH model using Wannier functions (see Fig.~\ref{MLWF_plot} and (\ref{wannier_psi_a2})-(\ref{wannier_psi_b2})). Here we compare this approach with ours. Specifically, we compare the 
interaction coefficients found via an expansion in terms of the MLWFs. To our surprise, they are quite close. 

To obtain 
coefficients from the direct Wannier approach, we use two-band maximally localized Wannier functions (MLWFs) computed using the eigenfunctions of the $PXP$ matrix defined in (\ref{project_op}) and (\ref{position_op}). We also compared these MLWFs with those obtained by the steepest descent algorithm given in \cite{Marzari1997}. Both methods gave essentially the same MLWFs. However,   the $PXP$ approach yields completely real functions, whereas the steepest descent algorithm required rescaling by a (complex) constant to impose reality. 


\begin{table}
\centering
    \begin{tabular}{||c|c|c||}
        \hline
        Optimization Coeffs & Wannier Coeffs & Absolute Diff\\
        \hline\hline
        $x_1$ = 11.1067848512 & 11.1066166640 & 1.681872$\times 10^{-4}$\\
        $x_2$ = 4.6407429151 & 4.32899283674 & 0.3117500785\\
        $x_3$ = 8.4987551661 & 8.65727412801 & 0.1585189619\\
        \hline
    \end{tabular}
\caption{Tight-binding coefficients obtained by the optimization approach and a direct MLWF expansion using $r = 3$ coefficients. The potential used to construct these values is (\ref{pot_expand}) with the parameters $V_0^2 = 100$, $\sigma = 0.05$, and potential peaks located at $-0.30,0.30$. \label{table_compare_wannier}}
\end{table}

To ensure our method converges to MLWF-type coefficients in an efficient manner, we employ the following protocol: first, run the program and find the three coefficient fit. Use the three coefficient solution as the initial guess for the six coefficient model. 
Run the program for six coefficients and repeat the process for nine coefficients using the six coefficient solution. This will allow the program to find interaction values that closely agree with the Wannier interaction coefficients for most tolerances. 

A comparison of self and nearest neighbor interactions for $r = 3$ is shown in Table~\ref{table_compare_wannier}.
 The self-interaction is remarkably accurate, agreeing to nearly four decimal places; $\mathcal{O}(10^{-5})$ relative error. The nearest neighbor coefficients  ($x_2,x_3$)  are found to agree to nearly one decimal place; $\mathcal{O}(10^{-2})$ relative error. Next, we increase the number of coefficient to $r = 6$; the data is shown in Table~\ref{compare_coeff_6}. Both self-interaction terms ($x_1,x_4$) exhibit good agreement; $\mathcal{O}(10^{-5})$ relative error. Remarkably, increasing the total number of interactions improves the agreement with the MLWFs by about a factor of 2.6-2.8 for both nearest neighbor coefficients.  Lastly, the next-nearest neighbor coefficients show similar level of agreement with their MLWF-derived counterparts.

We conjecture this is the reason for the close agreement between the continuous and discrete edge modes in Fig.~\ref{cont_disc_edge_modes}. Our optimization approach appears to be rapidly deriving the coefficient of the MLWF approach, but at a fraction of the time and at minimal inconvenience to the user. 

Unfortunately, the agreement between our coefficients and the MLWF coefficients does not improve  as we include more couplings. That is, for $r = 9$ coefficients the difference between the MLWF and optimization coefficients grows, not decrease. Indeed, we find that our residual values are slightly lower than those obtained from the MLWFs. This is a result that merits further study in a future work.

\begin{table}
\centering
    \begin{tabular}{||c|c|c||}
        \hline
        Optimization Coeffs & Wannier Coeffs & Absolute Diff\\
        \hline\hline
        $x_1$ = 11.1067836954 & 11.1066166640 & 1.6703137$\times 10^{-4}$\\
        $x_2$ = 4.2085335649 & 4.3289928367 & 0.1204592718\\
        $x_3$ = 8.7122805327 & 8.6572741280 & 0.0550064047\\
        $x_4$ = -0.9892268588 & -0.9892687269 & 4.18681$\times 10^{-5}$ \\ 
        $x_5$ = 0.1964093132 & 0.2119304077 & 0.0155210945\\
        $x_6$ = 0.444110563 & 0.34153407391 & 0.1025764891\\
        \hline
    \end{tabular}
\caption{Tight-binding coefficients obtained by the optimization approach and a direct MLWF expansion using $r = 6$ coefficients. The potential used to construct these values has the same parameters as in Table \ref{table_compare_wannier}. \label{compare_coeff_6}}
\end{table}


\section{Conclusion}
\label{conclude_sec}

Tight-binding models are a well-established and common method for approximating lattice waveguide systems, including topological insulators.
In this work we established a least squares optimization algorithm that finds optimal fits for  SSH-type systems from given spectral band data and desired topology.  A user-friendly code written in MATLAB is available to implement this approach. This method was found to be  capable of approximating the spectral data to arbitrary accuracy  when either the potential depths are increased or the number of interaction coefficients increases.

There does appear to be some agreement between the fitted coefficients and those obtained from a Wannier expansion, however we did not observe convergence as the number of interaction coefficients increased. This is could be a fruitful future direction. That is, can the algorithm be modified to reproduce {\it the} coefficients obtained from a MLWF expansion. If so, this could offer an alternate route from traditional approaches.

Lastly, we anticipate this approach  has the ability to approximate other topological insulator systems too. The key is finding the appropriate ansatz model, e.g.  extended SSH model. Following the protocol laid out by Slater and Koster, the coefficients can be fitted based on important properties of the system, e.g. bulk bands, topology, etc. Future directions include other one-dimensional models as well as higher-dimensional systems. \\

\noindent
{\bf Code Availability} All  MATLAB codes used to produce the results  in this paper are freely available on the author's website (justinthomascole.com) as well as the GitHub site \url{https://github.com/mnameika/Numerical-SSH-Code}. 

\vspace{0.1 in}

\noindent
{\bf Acknowledgements} This work was supported by AFOSR research grant FA9550-23-1-0105, the Undergraduate Research Academy, and  Committee on Research and Creative Works seed grant at UCCS.

\appendix

\section{The SSH Model}
\label{SSH_review_sec}

In this appendix the SSH model and its solutions are reviewed. Here, only self and nearest neighbor interactions are considered. This model serves as the motivation for the extended models utilized in this work. 

\begin{figure}
    \includegraphics[scale = 0.35]{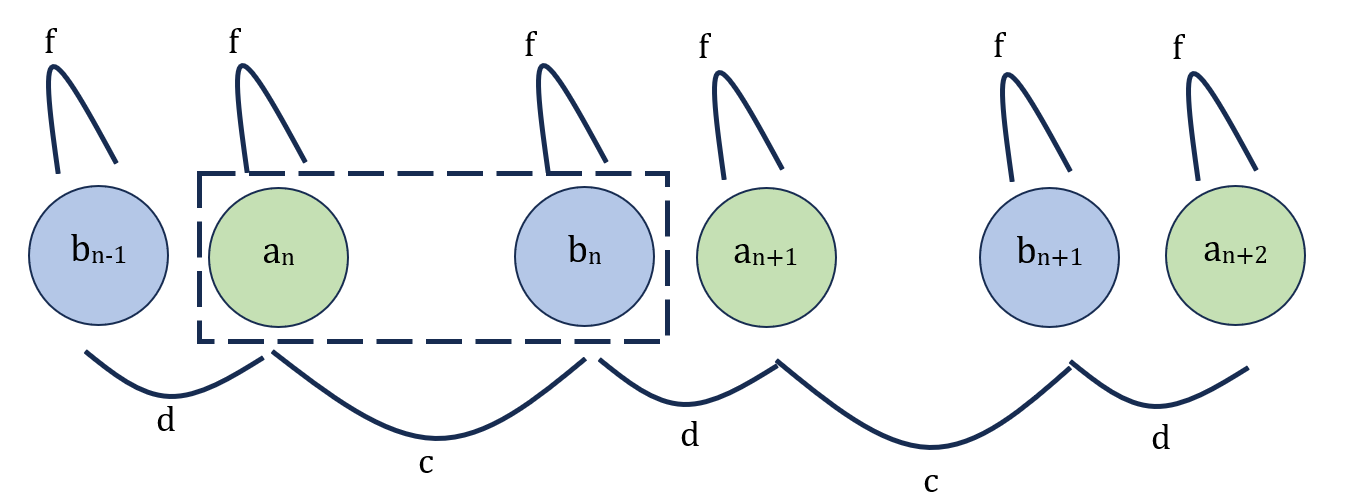}
    \centering
    \caption{Interactions among nearest neighbor sites (coefficients $c,d$) and self-interaction (coefficient $f$). The dashed rectangle depicts a typical unit cell. This physical configuration depicts a topologically nontrivial orientation corresponding to a nonzero winding number (\ref{zak_calc_SSH}).}
    \label{SSH_fig}
\end{figure}

Consider the classic SSH model 
\begin{equation}
\label{SSH}
    \begin{aligned}
        fa_n + cb_n + db_{n-1} &= \lambda a_n\\
        fb_n + ca_n + da_{n+1} &= \lambda b_n
    \end{aligned}
\end{equation}
 for $n \in \mathbb{Z}$, where $f $ is a self-interaction coefficient and $c, d $ are the nearest neighbor interactions (see Fig.~\ref{SSH_fig}). For a real potential $V(x)$ in (\ref{Schrodinger_eqn}), these coefficients are real. 

To begin, we exmaine bulk plane wave solution
 \begin{equation}
 \label{SSH_plane_waves}
        a_n = \alpha(k)e^{ikn}, ~~~~ b_n = \beta(k)e^{ikn} ,
 \end{equation}
 for wavenumber $k$. Doing so yields the following $2 \times 2$ bulk Hamiltonian
\begin{align}
    \label{SSH_bulk_eig_prob}
    {H}(k)  {\bm \gamma}(k)  &= \left[ \lambda(k) - f \right] {\bm \gamma}(k) , \nonumber \\
     {H}(k)  & =
     \begin{pmatrix}
        0 & c + de^{-ik}\\
        c + de^{ik} & 0
    \end{pmatrix}  ,
\end{align}
for eigenfunction ${\bm \gamma}(k) = (\alpha(k) , \beta (k))^T$.
Notice that spectrum has been shifted by an amount $f$ to remove the diagonal (on-site) terms. Doing this will center the spectral bands about  zero.
Observe that the bulk Hamiltonian possesses the following symmetries:
\begin{align*}
\sigma_x H(k) \sigma_x & =   H(-k)   , \\
\sigma_z H(k) \sigma_z & =  - H(k)  ,
\end{align*}
where $\sigma_x,\sigma_z$ denote the first and third Pauli matrices, respectively.
The first property implies that if $\lambda(k) - f$ is a nontrivial eigenvalue for a chosen $k$, then so is $\lambda(-k) - f$ . This shows an   {\it inversion} symmetry about $k = 0 $. The second property implies that if $\lambda(k) - f$ is a nontrivial eigenvalue for a given $k$, then so is $- (\lambda(k) - f)$. 
This is a {\it chiral} symmetry and indicates the existence of two distinct bands (equivalently sublattices).

\begin{figure}
\centering
    \includegraphics[scale = 0.23]{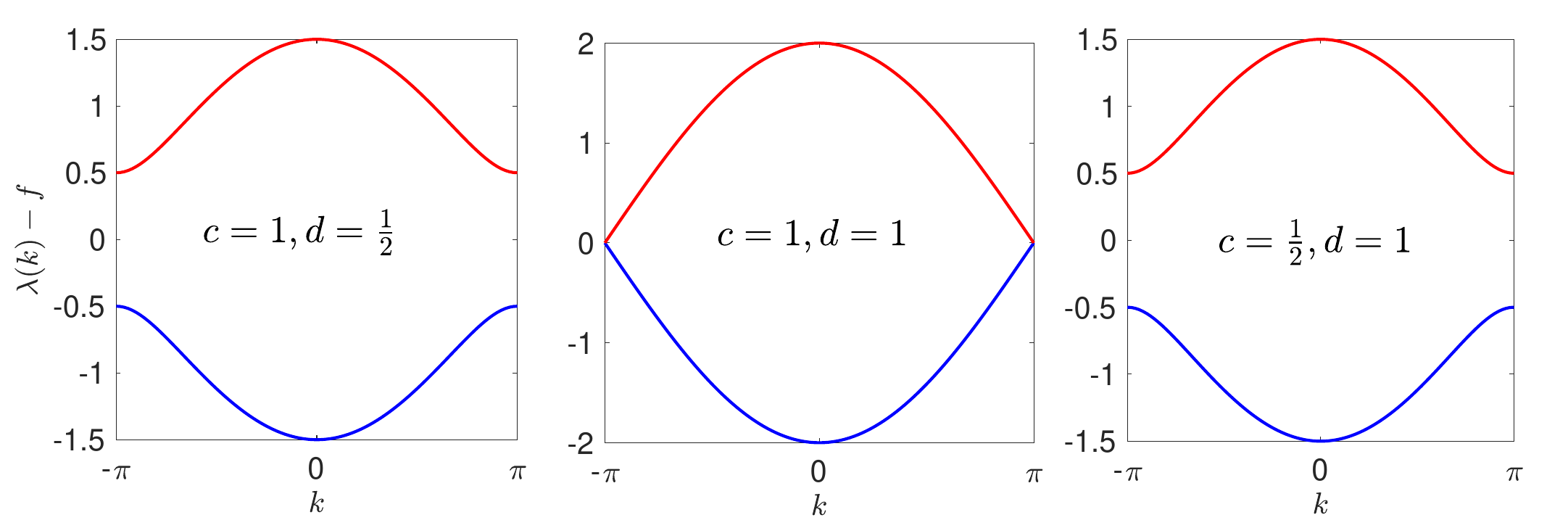}
    \caption{Bulk bands (\ref{SSH_bulk_bands_define}) for the various parameters. The top (bottom) band corresponds to $\lambda_+(k) - f$ $[\lambda_-(k) - f]$.}
    \label{SSH_bands_plot}
\end{figure}

The eigenvalues (bands) of (\ref{SSH_bulk_eig_prob}) are
\begin{equation}
\label{SSH_bulk_bands_define}
\begin{aligned}
    \lambda_1(k) &= f + |c + de^{ik}| = f + \sqrt{c^2 + 2 cd \cos (k) + d^2} , \\
    \lambda_2(k) &= f - |c + de^{ik}| = f - \sqrt{c^2 + 2 cd \cos(k) + d^2},
\end{aligned}
\end{equation}
where $\lambda_1(k)$ corresponds to the ``upper'' band, and $\lambda_2(k)$  the is  ``lower'' band. 
A couple of typical bands are shown in Fig.~\ref{SSH_bands_plot}.
Notice that these bands satisfy the inversion and chiral symmetries. Also notice that the band in  Fig.~\ref{SSH_bands_plot}(left) and Fig.~\ref{SSH_bands_plot}(right)  are identical since the functions in (\ref{SSH_bulk_bands_define}) are invariant under the symmetry $(c,d) \rightarrow (d,c)$, i.e. $\lambda$ is invariant under exchanging $c$ and $d$. The corresponding normalized eigenfunctions are
\begin{eqnarray}
    \label{SSH_eigenvector_define}
    {\bm \gamma}_{\pm}(k) &&= \frac{1}{\sqrt{2}}
    \begin{pmatrix}
    \pm \frac{c+ d e^{- i k}}{| c+ d e^{- i k}|} \\ 1
    \end{pmatrix}
     =  \frac{1}{\sqrt{2}}
     \begin{pmatrix}
     \pm e^{i \theta(k)} \\ 1
     \end{pmatrix} , \nonumber\\
     \theta (k) &&= \arg \left(  \frac{c+ d e^{- i k}}{| c+ d e^{- i k}|}\right) .
\end{eqnarray} 
Notice that the nontrivial first component $ \pm e^{i \theta(k)} $ lies on the unit circle. 
The topology of the function is revealed through these eigenfunctions. 

\begin{figure}
\centering
    \includegraphics[scale = 0.25]{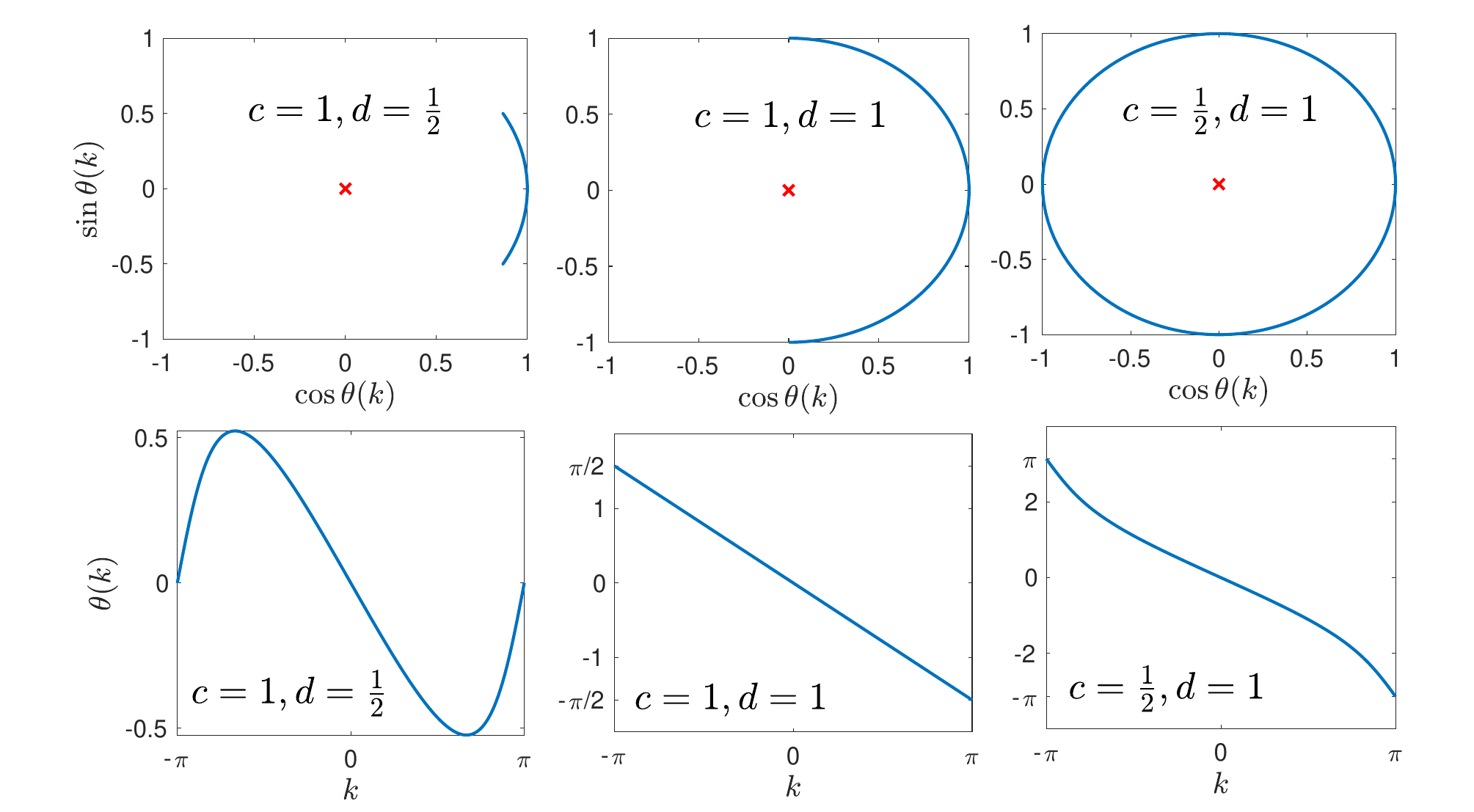}
    \caption{Top row: path of $\theta(k)$ defined in (\ref{SSH_eigenvector_define}) for the same coefficients used in Fig.~\ref{SSH_bands_plot}. Bottom row: angle (in radians) as a function of $k$ over the Brillouin zone $[-\pi, \pi]$. Use the angle in the bottom row to find the position on the unit circle in the top row. The Zak phase of the left (right) column is zero (one).}
    \label{SSH_winding_plot}
\end{figure}

For the eigenvector (\ref{SSH_eigenvector_define}), equipped with the complex dot product, the Zak phase is
\begin{eqnarray}
   \label{zak_calc_SSH}
    \mathcal{Z} &&= \oint_C i \left\langle   {\bm \gamma} (k) \bigg\rvert \frac{d {\bm \gamma} }{d k} \right\rangle dk  = -\frac{1}{2} \oint_C \frac{d \theta}{dk} dk \nonumber\\
    &&= - \frac{1}{2} \left[ \theta(k_{\rm final})  -\theta(k_{\rm initial})   \right] .
\end{eqnarray}
The Zak phase value (\ref{Zak_phase_cont}) corresponds to whether the path of $\theta(k)$ in (\ref{SSH_eigenvector_define}) encloses the origin, hence it is a winding number.
 The paths for  different topological cases are displayed in Fig.~\ref{SSH_winding_plot}. The bottom row shows the angle $\theta(k)$ as $k$ moves from $-\pi$ to $\pi$. Suppose $0 <  |d| < |c|$. Then the total change in angle $\Delta \theta = 0$ and does not enclose the origin (see Fig.~\ref{SSH_winding_plot}(left)). As a result, the Zak phase is trivial and
$    \mathcal{Z} = - \frac{1}{2\pi} \left[ 0 - 0  \right] = 0 .$
The case $ 0 < |c| = |d|$ is degenerate; this corresponds to a gap closure in Fig.~\ref{SSH_bands_plot}(middle) and a topological transition point (see Fig.~\ref{SSH_winding_plot}(middle)).  Lastly, suppose  $|d| > |c| > 0$. Then the entire unit circle is traversed in a clockwise fashion ($\Delta \theta = - 2 \pi $) as $ k$ moves from $- \pi$ to $  \pi$ (see Fig.~\ref{SSH_winding_plot}(right)). As a result, the Zak phase takes a nontrivial (topological) value
   $ \mathcal{Z} = - \frac{1}{2} \left[ -\pi - (\pi)  \right] = \pi .$

Let us next construct the edge problem which complements the bulk problem studied above. We consider a semi-infinite  boundary value problem such that $|a_n|,|b_n| \rightarrow 0$ as $n \rightarrow \infty$. This is the left edge mode. There is a similar right edge mode localized on the right boundary. To compute these modes, consider the SSH system (\ref{SSH}) with Dirichlet zero (open) boundary conditions
\begin{equation*}
    a_n = 0 ,~  b_n = 0  ~~~~~ {\rm for}~~ n < 1 ~{\rm and}~ n > N ,
\end{equation*}
where $N \gg 1$. The resulting $2N \times 2N$ block matrix system is

\begin{eqnarray}
\label{edge_eig_system}
\mathcal{H} {\bm v} &&= \lambda {\bm v}, ~~~~
   \mathcal{H} = \begin{bmatrix}
        \mathcal{H}_{11} & \mathcal{H}_{12}\\
        \mathcal{H}_{21} & \mathcal{H}_{22}
    \end{bmatrix}, \nonumber \\
  {\bm v} &&= \left( a_1, a_2, \dots, a_N | b_1, b_2, \dots, b_N  \right)^T  ,
\end{eqnarray}
for the $N \times N$ matrices 

\begin{eqnarray}
    \mathcal{H}_{11} &&= \begin{pmatrix}
        f & & & \\
        & f & & \\
         & & \ddots & \\
         & & & f
    \end{pmatrix}, \hspace{1em} \mathcal{H}_{12} = \begin{pmatrix}
        c & & & \\
        d & c & & \\
        & \ddots & \ddots & \\
        & & d & c
    \end{pmatrix}, \nonumber \\
    \hspace{1em} \mathcal{H}_{21} &&= \mathcal{H}_{12}^T, \hspace{1em} \mathcal{H}_{22} = \mathcal{H}_{11} .
\end{eqnarray}

\begin{figure}
\centering
    \includegraphics[scale = 0.24]{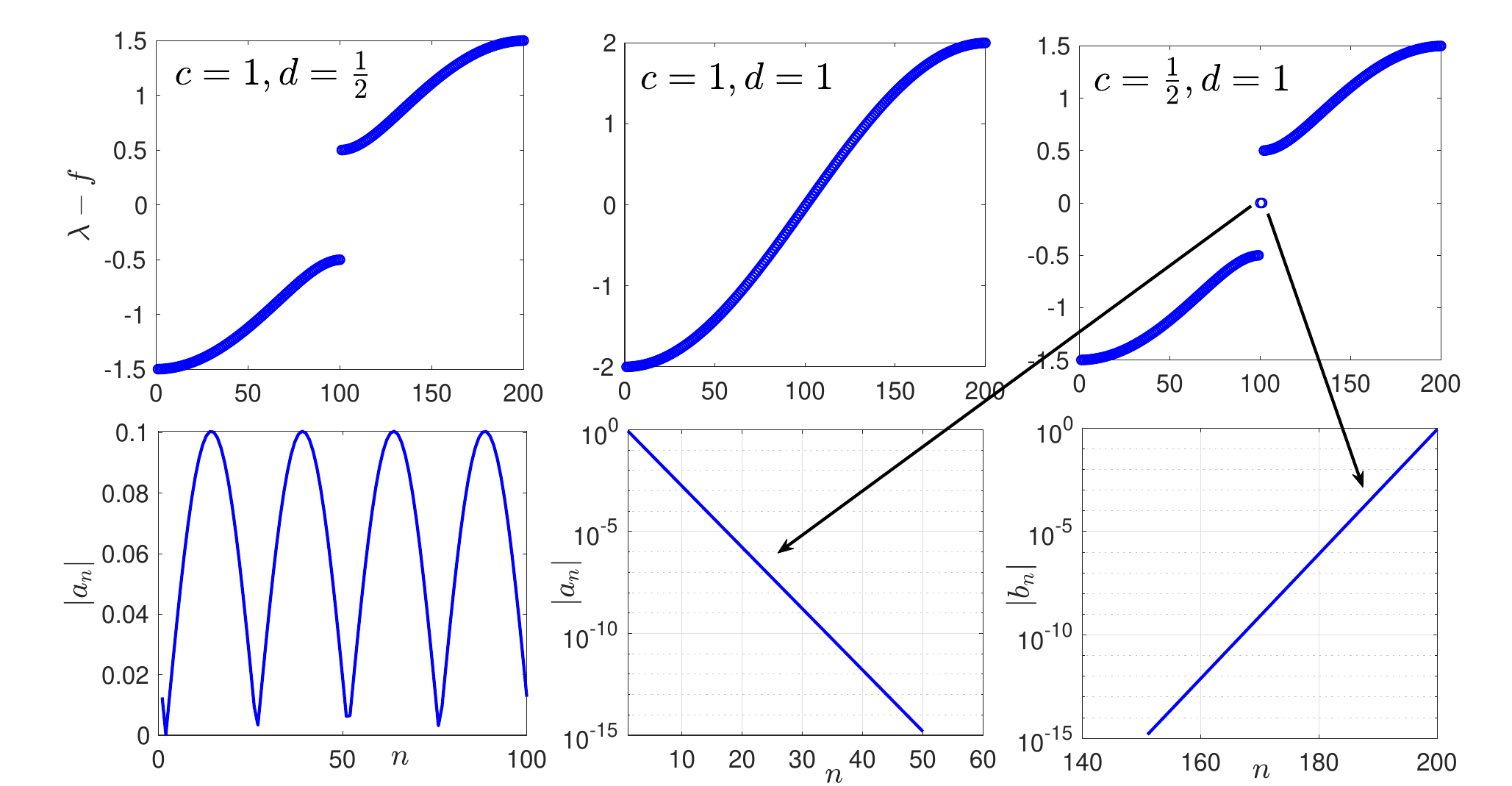}
    \caption{Top row: All $2N$ Eigenvalues (sorted in ascending order, $N = 100$) computed from  (\ref{edge_eig_system}) using the same coefficients as in Fig.~\ref{SSH_bands_plot}. 
    When $|c| < |d|$ there are two isolated eigenvalues at $\lambda = f$ (see top right panel). 
    Bottom row:  eigenmodes from the topologically nontrivial problem ($c= 0.5, d = 1$). 
    From left-to-right, a typical bulk mode ($\lambda \not= f$) and the localized edge eigenmodes ($\lambda = f$) are shown. 
    The magnitude of the edge modes is plotted on a logarithmic scale to highlight their exponential decay rate (see (\ref{SSHedge_solutions})).}
    \label{edge_spec_modes_plot}
\end{figure}

A typical set of results found by solving eigenvalue problem (\ref{edge_eig_system}) is shown in Fig.~\ref{edge_spec_modes_plot}. 
In the top row the eigenvalues, sorted in ascending order, are shown.  The bottom row of Fig.~\ref{edge_spec_modes_plot} shows the magnitude of the eigenmodes: a typical bulk mode and the localized modes. 
In the trivial case ($\mathcal{Z} = 0$) all eigenmodes are extended, bulk functions (similar to bottom left panel). 
In the nontrivial case ($\mathcal{Z} = \pi$) there exists two localized eigenmodes at $\lambda = f$. 
Notice that both edge modes decay exponentially fast away from the endpoints, as indicated by the linear form in the semi-log plots. On the left (right) edge, the modes $b_n = 0$ ($a_n = 0$) and are not shown.

Let us examine these localized edge modes closer. Take $\lambda  = f$ in (\ref{SSH}) which decouples the two difference equations and gives
\begin{equation}
    \begin{aligned}
        cb_n + db_{n-1} &&= 0 ,  \\
        ca_n + da_{n+1} &&= 0 .
    \end{aligned}
\end{equation}
These equations reveal the solution structure
\begin{equation}
\frac{a_{n+1}}{a_{n}} = - \frac{c}{d}  ~~~ {\rm and} ~~~  \frac{b_n}{b_{n-1}} = - \frac{d}{c},
\end{equation}
whose right-hand sides are reciprocals of each other, 
and exponential solutions
\begin{equation}
    \label{SSHedge_solutions}
    a_n = A \left( - \frac{c}{d} \right)^{n}, ~~ b_n = B \left( - \frac{d}{c} \right)^{n}, ~~~ n  = 1, 2, \dots , N
\end{equation}
for  constants $A,B$. 
Consider the topological case when $| c/ d| < 1$. As $N \rightarrow \infty$ only  $a_n$ decays to zero. 
In order to satisfy the right boundary conditions one must set $B = 0$ and $b_n = 0$. On the other end, as $n \rightarrow 1$ for $N \gg 1$ the $a_n$ grows unbounded and so $A = 0$ and $a_n = 0$.  This describes the  two (distinct) eigenmodes shown in the topological case in Fig.~\ref{edge_spec_modes_plot}.

\section{Numerical approximation of the Zak phase}
\label{Zak_appendix}

In this section we discuss the numerical approximation of the discrete Zak phase. For the continuous problem, we compute the Zak phase using MLWFs. To begin, discretize the Brillouin zone interval $\left[ - \frac{\pi}{L}, \frac{\pi}{L} \right]$ with periodic boundary conditions. 
Notice  the Berry connection boundary condition 
$ \mathcal{A}\left( k + \frac{2 \pi }{L} \right)   = \mathcal{A}\left( k \right) .$ 
Consider  the equally spaced discretized $k$-space $k_j = - \frac{\pi}{L} + (j-1) \Delta k$ for $j = 1, 2, \dots, M+1$, where $\Delta k = \frac{2 \pi }{L M} $. The discretized periodic boundary conditions are $v_{j + M  } = v _ j$. Hence, $v_{M+1} = v_1$.

At each value $k_j$, we numerically compute the eigenvectors ${\bf v}_j \equiv {\bf v}(k_j)$ of the associated bulk Hamiltonian matrix.  To approximate the Zak phase (\ref{Zak_phase_cont}), first the integrand is approximated by 
$$\langle {\bf v}(k) | {\bf v}(k + \Delta k)   \rangle  \approx  1 + \Delta k \left\langle {\bf v}(k) \bigg|  \frac{d {\bf v}}{ d k} (k) \right\rangle ,$$
where we fix 
$\langle {\bf v}(k) | {\bf v}(k)   \rangle  = 1$. 
This is a linear approximation valid when $0 <\Delta k \ll 1$. Note that the second term (up to a factor of $\Delta k$) represents the instantaneous geometric phase $i \theta(k)$ and is intrinsically an imaginary function. Also, note that the logarithm of this quantity above yields
\begin{align}
\ln \left[  \langle {\bf v}(k) | {\bf v}(k + \Delta k)   \rangle  \right]  &\approx \ln \left[  1 + \Delta k \left\langle {\bf v}(k) \bigg| \frac{d {\bf v}}{ d k} (k) \right\rangle  \right] \nonumber \\
&\approx  \Delta k \left\langle {\bf v}(k) \bigg|  \frac{d {\bf v}}{ d k} (k) \right\rangle  ,
\end{align}
using the Taylor series for $\ln (1+ x)$ when $|x| \ll 1$.

The Zak phase (\ref{Zak_phase_cont}) contour integral is  approximated by the trapezoidal rule on a periodic domain
\begin{eqnarray}
\mathcal{Z} &&=  i  \oint \left\langle {\bf v} \bigg| \frac{d {\bf v} }{ d k} \right\rangle dk \nonumber \\
&& \approx   i  \sum_{j= 1}^M   \ln \left[  \langle {\bf v}_j | {\bf v}_{j+1}  \rangle  \right] \nonumber \\
&& =  -  \sum_{j= 1}^M \text{Im} \ln \left[  \langle   {\bf v}_j | {\bf v}_{j+1}  \rangle  \right] .
\end{eqnarray}
This is the manner in which the Zak phase is numerically approximated.

Define the $M \times 1$ instantaneous phase change vector ${\bm \gamma}$
\begin{equation}
    {\bm \gamma} = \begin{pmatrix}
        \langle {\bf v}_1 | {\bf v}_{2} \rangle \\
        \langle {\bf v}_2 | {\bf v}_{3} \rangle \\
        \vdots\\
        \langle {\bf v}_{M-1} | {\bf v}_M \rangle \\
        \langle {\bf v}_M | {\bf v}_1 \rangle
    \end{pmatrix} , 
\end{equation}
where the periodicity of ${\bf v}_j$ has been applied in the last element.
 The natural inner product is taken. 
 
 The Zak phase is approximated  by summing the negative imaginary parts of each element of ${\bm \gamma}$, i.e.
\begin{equation}
\label{linear_approx_zak}
    \mathcal{Z} = - \sum_{j=1}^M \text{Im}({\ln({\bm \gamma_j})}),
\end{equation}
where ${\bm \gamma}_j$ denotes the $j^{\rm th}$ element of ${\bm \gamma}$.

\section{Numerical computation of Wannier modes}
\label{num_calc_wannier_modes}

In this section we discuss the numerical computation of Wannier modes used in this work. This is relevant for Sec.~\ref{wannier_comparison} where we compare the coefficients of our approach to that of a traditional Wannier expansion.
A popular approach to computing well localized Wannier functions is the maximally localized Wannier function (MLWF) algorithm \cite{Marzari1997}.  This established technique 
 chooses a smooth gauge that minimizes the spread or width of the Wannier modes. Said differently, this algorithm searches for the most localized Wannier modes. 

An alternative approach, that we find more effective for this problem, is the eigenfunctions of the $PXP$ operator. 
 The two-band projection operator $P$ (not to be confused with the inversion operator $\mathcal{P}$), is defined by \cite{Vanderbilt2018}
\begin{eqnarray}
    \label{projection_op}
    P &&=\frac{L}{2 \pi}  \int_{\rm BZ}    \sum_{p = 1}^2   | \psi_p(x,k) \left> \right< \psi_p(x,k) |  ~ dk \nonumber\\
    &&\approx \sum_k   \sum_{p = 1}^2   | \psi_p(x,k) \left> \right< \psi_p(x,k) |  ,
\end{eqnarray}
where the integral/sum in $k$ is over the Brillouin zone.
These inner products are chosen to satisfy orthonormality condition $\langle \psi_m | \psi_n \rangle = \delta_{mn}$, where $m,n$ denote band indices. 

The natural inner product here is $\langle f | g \rangle = \int_{- \ell/2}^{\ell/2}  f^*(x) g( x) dx$. Consider discretizing the interval $\left[ - \frac{\ell}{2} ,  \frac{\ell}{2} \right]$, of length $ \ell$, by $x_j = - \frac{\ell}{2}  + j \Delta x$ for $j = 0, 1,  \dots, N$ and grid spacing $\Delta x =  \ell / N$.  Note that $\ell$ is not necessarily $L$, the periodic interval. Below, we only extend an integer amount the periodic interval, that is $\ell = \eta L$, where $\eta$ is some positive integer.
Define the discretized (in space) Bloch wave vector
$$| \psi_p (x, k ) \rangle  = {\bm \psi}_p(k) =   (\psi_p(x_0,k) , \psi_p(x_1,k), \dots, \psi_p(x_N,k)  )^T ,$$
where $p$ denotes the band number. If the spatial domain is longer than the unit cell $[-L/2 , L/2]$, the quasi-periodic boundary conditions are used to extend it, namely $\psi(x - L , k) = e^{- i k  L} \psi(x,k) $.
As such, the  explicit representation of the projection operator is 
\begin{equation}
    \label{project_op}
    P = \sum_k {\bm \psi}_1 {\bm \psi}_1^\dag  +  {\bm \psi}_2 {\bm \psi}_2^\dag ,
\end{equation}
where $^\dag$ denotes the complex conjugate transpose. 

Next, $X$ denotes the position operator. This matrix is diagonal and contains the locations of the spatial grid points, i.e.
\begin{equation}
    \label{position_op}
    X = \text{diag}\left( x_0 , x_1, \dots, x_N \right) ,
\end{equation}
where $X$ is a square matrix with dimensions $(N + 1) \times (N + 1)$. The $PXP$ is the product of these three matrices, in that order. The eigenfunctions of the $PXP$ operator are the MLWFs \cite{Vanderbilt2018}. In our experience, the Wannier eigenfunctions were always found to be completely real. Moreover, they are found match the multiband Wannier functions obtained iteratively in \cite{Marzari1997}, up to a (complex) scalar. Indeed one can show that the projection operator (\ref{projection_op}) is invariant under unitary transformations. Lastly, we note that the Wannier functions shown in Fig.~\ref{MLWF_plot} are for single band; either $p = 1$ or $p =2$, but not both.

\bibliography{datadrivenbib}

\end{document}